\documentclass[12pt]{article}
\usepackage{putex}
\usepackage{graphicx}

\begin{document}

\preprint{PUPT-2309}

\institution{PU}{Joseph Henry Laboratories, Princeton University, Princeton, NJ 08544}

\title{Ground states of holographic superconductors}

\authors{Steven S. Gubser and Abhinav Nellore}

\abstract{We investigate the ground states of the Abelian Higgs model in $AdS_4$ with various choices of parameters, and with no deformations in the ultraviolet other than a chemical potential for the electric charge under the Abelian gauge field.  For W-shaped potentials with symmetry-breaking minima, an analysis of infrared asymptotics suggests that the ground state has emergent conformal symmetry in the infrared when the charge of the complex scalar is large enough.  But when this charge is too small, the likeliest ground state has Lifshitz-like scaling in the infrared.  For positive mass quadratic potentials, Lifshitz-like scaling is the only possible infrared behavior for constant nonzero values of the scalar.  The approach to Lifshitz-like scaling is shown in many cases to be oscillatory.}

\date{August 2009}

\maketitle

\tableofcontents

\section{Introduction}

The Abelian Higgs model in $AdS_4$ is specified by the action
 \eqn{Lagrangian}{
  S = {1 \over 2\kappa^2} \int d^4 x \, \sqrt{-g}\left[ R - 
    {1 \over 4} F_{\mu\nu}^2 - |(\partial_\mu - i q A_\mu) \psi|^2 - 
    V(\psi,\psi^*) \right] \,,
 }
where $V$ is assumed to depend on $\psi$ and $\psi^*$ only through the product $\psi\psi^*$.  The theory \eno{Lagrangian} was introduced in \cite{Gubser:2008px} with the aim of describing superconducting black holes, following earlier work \cite{Gubser:2005ih} on black hole phase transitions and \cite{Herzog:2007ij,Hartnoll:2007ih,Hartnoll:2007ip}, among others, on the possible relation between $AdS_4$ vacua and quantum critical behavior.  In \cite{Hartnoll:2008vx}, the Abelian Higgs model was treated in a probe approximation, where the matter fields do not back-react appreciably on the metric.  This approximation is justified in the limit of large $q$, and it is a useful starting point for studies of the conductivity and the behavior near the phase transition.  However, one must go beyond the probe approximation to discover what the energetically preferred zero-temperature states are.  In \cite{Gubser:2008wz}, it was suggested that for the W-shaped quartic potential
 \eqn{VChoice}{
  V(\psi,\psi^*) = -{6 \over L^2} + m^2 \psi\psi^* + 
    {u \over 2} (\psi\psi^*)^2 \,,
 }
with $m^2 < 0$ and $u>0$, the zero-temperature limit of superconducting black holes is a domain wall that interpolates between the $AdS_4$ vacuum with $\psi=0$ and the symmetry-breaking $AdS_4$ vacuum with
 \eqn{psiIR}{
  |\psi| = \psi_{\rm IR} \equiv \sqrt{-m^2 \over u} \,.
 }
An example of such a domain wall was exhibited in \cite{Gubser:2008wz}, but it was left open whether such a domain wall exists for all values of $q$, and whether it really is a zero-temperature limit of superconducting black holes.  The next studies beyond the probe approximation were \cite{Hartnoll:2008kx,Gubser:2008pf}.  These studies focused on the case $m^2 < 0$, $u=0$, and they provided numerical evidence that in the $T \to 0$ limit, all the charge is expelled from the black hole---at least when $q$ is not too small and/or $m^2 L^2$ is sufficiently negative.  Moreover, according to \cite{Gubser:2008pf}, an $SO(2,1)$ boost symmetry appears in the infrared of the $T=0$ solution, similar to the conformal symmetry of the infrared side of the domain walls studied in \cite{Gubser:2008wz}.  Subsequently, it was shown in \cite{Gubser:2009gp} that $AdS_5$-to-$AdS_5$ and $AdS_4$-to-$AdS_4$ domain walls exist in string theory and M-theory, based on theories similar to \eno{Lagrangian} but with curved target spaces for the scalars.  The $AdS_5$-to-$AdS_5$ case was based on \cite{Gubser:2009qm} and was reported on earlier in \cite{GubserStrings}, and the $AdS_4$-to-$AdS_4$ case was previously suggested in \cite{Gauntlett:2009dn}, based in part on \cite{Gauntlett:2009zw}.

In a parallel line of development, it was demonstrated in \cite{Kachru:2008yh} that a massive gauge field coupled to gravity leads to geometries with anisotropic, Lifshitz-like scaling: $t \to \lambda^z t$ while $\vec{x} \to \lambda \vec{x}$ for some critical exponent $z$.  It was shown in \cite{Azeyanagi:2009pr} that solutions to type~IIB supergravity exist with anisotropic scaling between different spatial dimensions; moreover, such scaling solutions could be obtained as the infrared limit of flows from a suitably deformed $AdS_5$ vacuum.  Some no-go arguments were given in \cite{Li:2009pf} against the existence of Lifshitz solutions in type~IIA supergravity and M-theory.

The problem of finding superconducting black hole solutions to the classical equations of motion following from \eno{Lagrangian} can be posed as follows.  Consider the ansatz
 \eqn[c]{AdSAnsatz}{
  ds^2 = e^{2A(r)} 
    \left[ -h(r) dt^2 + d\vec{x}^2 \right] + e^{2B(r)}{dr^2 \over h(r)}  \cr
  A_\mu dx^\mu = \Phi(r) dt \qquad \psi = \psi(r) \qquad B=0\,,
 }
where $\psi(r)$ is everywhere real, and $B=0$ is a gauge choice.  The equations of motion and zero-energy constraint are
 \begin{eqnarray}
  A'' &=& -{1 \over 2} \psi'^2 - 
    {q^2 \over 2h^2 e^{2A}} \Phi^2 \psi^2  \label{Aeom} \\
  h'' + 3 A' h' &=& e^{-2A} \Phi'^2 + 
    {2q^2 \over h e^{2A}} \Phi^2 \psi^2  \label{heom}
 \end{eqnarray}
 \begin{eqnarray}
  \Phi'' + A' \Phi' &=& {2q^2 \over h} \Phi \psi^2  \label{Phieom} \\
  \psi'' + \left( 3A' + {h' \over h} \right) \psi' &=& 
    {1 \over h} {\partial V \over \partial\psi^*} - 
      {q^2 \over h^2 e^{2A}} \Phi^2 \psi \label{psieom}
 \end{eqnarray}
 \eqn{ZeroEnergy}{
    h^2 \psi'^2 + e^{-2A} q^2 \Phi^2 \psi^2 -
   {1 \over 2} h e^{-2A} \Phi'^2 - 2 hh' A'  
   - 6 h^2 A'^2 -
   h V(\psi,\psi^*) = 0 \,.
 }
It is straightforward to show that
 \eqn{NoetherCharge}{
  Q = e^A(e^{2A}h'-\Phi\Phi')
 }
is a constant if the equations of motion \eno{Aeom}-\eno{ZeroEnergy} are satisfied.  It is the Noether charge associated with the scaling symmetry
\eqn[c]{NoetherSymmetry}{
  A \to A-\log c \qquad h \to c^{6} h \qquad \Phi \to c^2\Phi \qquad B \to {B+3\log c}  \cr
  t \to t/c^2 \qquad \vec{x} \to c \vec{x}
}
of the action \eno{Lagrangian} when evaluated with $A_\mu$ and the metric as in \eno{AdSAnsatz} before $B$ has been fixed. If there is a black hole horizon at $r=r_H$, then the temperature and entropy density are
\eqn{Tands}{
 T={e^{A(r_H)}h'(r_H)\over 4\pi} \qquad\qquad s={2\pi\over\kappa^2} e^{2A(r_H)}
  \,,
}
and one sees that
 \eqn{Noether}{
  Q=2\kappa^2 T s \,.
 }
The $\Phi\Phi'$ term drops out of the relation \eno{Noether} because $\Phi(r_H)$ has to be zero in order for $\Phi dt$ to be well-defined at the horizon as a one-form, and $\Phi'(r_H)$ has to be finite so as to avoid generating divergent stress-energy.  Thus $Q=0$ is a form of extremality condition: It implies that either there is no horizon at all, or that if there is one, it has $Ts=0$.

The behavior of the fields near the conformal boundary of $AdS_4$ is
 \eqn{FarBCs}{
  A &= \sqrt{H_0}{r \over L} + a_0 + \ldots  \cr
  h &= H_0 + H_3 e^{-3A} + \ldots  \cr
  \Phi &= p_0 + p_1 e^{-A} + \ldots   \cr
  \psi &= \Psi_a e^{(\Delta_\psi-3)A} + \Psi_b e^{-\Delta_\psi A}
    + \ldots \,,
 }
where $\ldots$ stands for terms that are subleading at large $r$ relative to the ones shown, and
 \eqn{DeltaPsi}{
  \Delta_\psi (\Delta_\psi-3) = m^2 L^2 \,.
 }
(We will restrict attention to the larger root of this equation even in the window where both roots correspond to valid operator dimensions.) The chemical potential $\mu$, the charge density $\rho$, and the energy density $\epsilon$ of the dual gauge theory are obtained from asymptotics near the boundary as
 \eqn{ThermoEqs}{
 \mu &= {p_0\over 2L\sqrt{H_0}} \qquad
 \rho = -{p_1\over \kappa^2\sqrt{H_0}} \qquad
 \epsilon = -{H_3\over\kappa^2 L H_0} \,.
 } Consider fixing $p_0$ at some definite value and setting $a_0=0$, $H_0=1$, and $\Psi_a = 0$.  This corresponds to studying the dual gauge theory at finite chemical potential but not deforming its lagrangian with the operator ${\cal O}_\psi$ dual to $\psi$.  Alternatively, one may leave $p_0$ free and instead fix $p_1$: This corresponds to considering the dual gauge theory at fixed charge density.  Evaluating the conserved charge \eno{Noether} close to the boundary gives
\eqn{ConformalThermo}{
\epsilon={2\over 3}(Ts + \mu\rho)\,.
}
This relationship also follows from the tracelessness of the field theory stress-energy tensor, which implies
\eqn{TracelessT}{
\epsilon-2p=0\,.
}
Above, $p$ is the pressure. For a large, homogeneous system at finite temperature and chemical potential, the pressure is just $-g$, where the Gibbs free energy density $g$ is defined as
\eqn{GibbsDef}{
g=\epsilon-Ts-\mu\rho\,.
}
Equations \eno{TracelessT} and \eno{GibbsDef} together imply \eno{ConformalThermo}.  The conserved charge \eno{Noether} thus enforces a thermodynamic relationship that holds for the dual conformal theory by connecting bulk thermodynamic variables that appear in horizon and boundary asymptotics.

Solving the equations \eno{Aeom}-\eno{ZeroEnergy} with the boundary conditions described in the previous paragraph, and demanding no singularities in the bulk outside regular black hole horizons, one might find only the $AdS_4$ Reissner-Nordstrom black hole solution (hereafter RNAdS), where $\psi=0$ identically; or one may find superconducting solutions, where $\psi \neq 0$ spontaneously breaks the Abelian gauge symmetry.  Typically there are several one-parameter families of solutions, each one parametrized by the energy density.  The question at issue is what happens when we make this energy density as small as possible.  In other words, what is the ground state of the system at finite chemical potential, or at finite charge density?  If superconducting black holes are stable and thermodynamically favored over RNAdS, then this question is the same as asking what the zero-temperature limit of superconducting black holes is.

A reasonable guess is that when $m^2 < 0$ and $u>0$, the zero-temperature limit is always a domain wall like the one in \cite{Gubser:2008wz}, with emergent conformal symmetry in the infrared.  By considering expansions around the infrared $AdS_4$ geometry, we will show in section~\ref{CONFORMAL} that this cannot be right when $q$ is too small.  We propose in section~\ref{LIFSHITZ} that what happens instead, below a certain threshold for $q$, is that the infrared geometry exhibits Lifshitz-like scaling.  This transition to Lifshitz behavior can be understood from a field theory perspective in terms of a non-conserved current operator becoming relevant when $q$ is below its threshold value.  We also find Lifshitz behavior when $m^2 > 0$.  We exhibit explicit, numerically generated examples of $AdS_4$-to-$AdS_4$ and $AdS_4$-to-Lifshitz domain walls.  All our analysis is based on simple four-dimensional gravity theories, not drawn from explicit string theory or M-theory constructions.

\section{Emergent conformal symmetry}
\label{CONFORMAL}

Let's assume that $V$ takes the simple quartic form \eno{VChoice}.  The existence of an $AdS_4$ vacuum with $\psi = \psi_{\rm IR}$ and $\Phi=0$ is wholly insensitive to the gauge field dynamics.  It is likely that one can flow to this vacuum from the $\psi=0$ $AdS_4$ vacuum with the gauge field set uniformly to $0$.  Such a holographic renormalization group flow, however, would have to be triggered by a relevant deformation of the lagrangian of the conformal field theory dual to the $\psi=0$ vacuum.  We are interested in eliminating such a deformation in favor of a finite density of the charge dual to the gauge field.  To inquire whether conformal symmetry can emerge in the infrared in this context, we must ask whether one can perturb the $\psi=\psi_{\rm IR}$ $AdS_4$ vacuum in such a way that it can match onto a domain wall solution with nonzero gauge field.  Let's express the $AdS_4$ vacuum as
 \eqn{IRvac}{
  ds^2 = e^{2r/L_{\rm IR}} (-dt^2 + d\vec{x}^2) + dr^2 \,,
 }
where
 \eqn{LIRdef}{
  L_{\rm IR} = \sqrt{-6 \over V(\psi_{\rm IR},\psi_{\rm IR})} \,.
 }
Then the perturbations of interest are ones that vanish in the $r \to -\infty$ limit (the deep infrared) and are either finite or divergent in the $r \to +\infty$ limit, which is eventually replaced by the domain wall.  In field theory terms, we wish to study irrelevant perturbations by operators dual to the fields $A_0$ and $\psi$.

As a first step, consider the linearized equations of motion for the scalar and the gauge field, assuming that $A_0=\delta\Phi$ is the only nonvanishing component of $A_\mu$, that the scalar $\psi = \psi_{\rm IR} + \delta\psi$ is everywhere real, and that both $\delta\Phi$ and $\delta\psi$ depend only on $r$:
 \eqn{LinIR}{
  \left[ \partial_r^2 + {1 \over L_{\rm IR}} \partial_r - 
     m_\Phi^2 \right] \delta\Phi &= 0  \cr
  \left[ \partial_r^2 + {3 \over L_{\rm IR}} \partial_r - 
     m_{\rm IR}^2 \right] \delta\psi &= 0 \,,
 }
where
 \eqn{mIRdefs}{
  m_\Phi^2 &= 2 q^2 \psi_{\rm IR}^2  \cr
  m_{\rm IR}^2 &= 2 {\partial^2 V \over \partial\psi \partial\psi^*}
   (\psi_{\rm IR},\psi_{\rm IR}) \,.
 }
Let us further define
 \eqn{DeltaIRdefs}{
  \Delta_\Phi &= {3 \over 2} + 
    \sqrt{{1 \over 4} + m_\Phi^2 L_{\rm IR}^2}  \cr
  \Delta_{\rm IR} &= {3 \over 2} + 
    \sqrt{{9 \over 4} + m_{\rm IR}^2 L_{\rm IR}^2} \,,
 }
where the positive sign on the square root is understood in both cases.  The operators $J^{\rm IR}_\mu$ and ${\cal O}_{\rm IR}$ dual to $A_\mu$ and $\delta\psi$ have dimensions $\Delta_\Phi$ and $\Delta_{\rm IR}$, respectively.  The solutions to \eno{LinIR} that vanish in the limit $r \to -\infty$ are
 \eqn{AllowedSolns}{
  \delta\Phi &\equiv \Phi_1 = 
     a_\Phi e^{(\Delta_\Phi-2) r/L_{\rm IR}}  \cr
  \delta\psi &\equiv \psi_1 =
     a_\psi e^{(\Delta_{\rm IR}-3) r/L_{\rm IR}} \,,
 }
where $a_\psi$ and $a_\Phi$ are undetermined coefficients.

The second formula in \eno{DeltaIRdefs} shows that $\Delta_{\rm IR} \geq 3$ provided $m_{\rm IR}^2 \geq 0$, which has to be true given that $\psi_{\rm IR}$ is a minimum of the potential.  In other words, the operator dual to $\psi$ at the infrared fixed point is an irrelevant perturbation, which makes sense because it participates in a flow toward conformality in the infrared.  The first formula in \eno{DeltaIRdefs} shows that $\Delta_\Phi \geq 2$ provided $m_\Phi^2 > 0$, which has to be true given the expression for $m_\Phi^2$ in \eno{mIRdefs}.\footnote{Also, there is a unitarity bound $\Delta_\Phi \geq 2$ for gauge-invariant, primary operators \cite{Mack:1975je} (see also \cite{Grinstein:2008qk}), suggesting that even in a more general setup, one cannot have $m_\Phi^2 < 0$.}  If $\Delta_\Phi > 3$, then the operator $J_0$ dual to $\Phi$ is also an irrelevant perturbation, so again one has a sensible field theory interpretation that $J_0$ participates in a flow toward Lorentz-invariant conformality in the infrared.  On the other hand, if $2 < \Delta_\Phi < 3$, then there seems to be a puzzle: In gravity we have the solution $\Phi_1$ exhibited in \eno{AllowedSolns}, which vanishes in the limit $r \to -\infty$; but in field theory, $J_0$ is a {\it relevant} operator, which should distort the field theory further and further away from Lorentz invariance as one proceeds toward the infrared.

The resolution of this puzzle is that gravity solutions describing charged matter in the ultraviolet conformal field theory cannot flow to the symmetry-breaking infrared fixed point if $2 < \Delta_\Phi < 3$.  As far as we can tell, this is the only obstacle to the existence of such flows.  This line of thought is what led to the Criticality Pairing Conjecture of \cite{Gubser:2009gp}.

To demonstrate the claim that flowing to a conformal fixed point is impossible (or at least fine-tuned) if $2 < \Delta_\Phi < 3$, we need to develop some machinery describing perturbations of the infrared conformal point.  Although the presentation of the next couple of paragraphs is a bit lengthy, the final punch line can be stated in advance: For $2 < \Delta_\Phi < 3$, there is strong back-reaction on the metric such that the blackening function $-g_{tt} / g_{xx}$, doesn't approach a constant in the infrared.  Instead, as we will describe in section~\ref{LIFSHITZ}, one finds Lifshitz-like scaling in the infrared.

Consider the expansions
 \eqn{FormalExpansions}{
  A &= {r \over L_{\rm IR}} + \lambda A_1 + \lambda^2 A_2 + 
    \lambda^3 A_3 + \ldots  \cr
  h &= 1 + \lambda h_1 + \lambda^2 h_2 + \lambda^3 h_3 + \ldots  \cr
  \Phi &= \lambda \Phi_1 + \lambda^2 \Phi_2 + \lambda^3 \Phi_3 + 
    \ldots  \cr
  \psi &= \psi_{\rm IR} + \lambda \psi_1 + \lambda^2 \psi_2 + \lambda^3 \psi_3 +
    \ldots \,,
 }
where $\lambda$ is a formal expansion parameter that we eventually want to set to unity.  What we are really expanding in is the smallness of all corrections to $AdS_4$ in the limit $r \to -\infty$.  Plugging the expansions \eno{FormalExpansions} into the equations of motion \eno{Aeom}-\eno{psieom}, one obtains at $n$th order in $\lambda$ the conditions
 \eqn{nthConditions}{
  \partial_r^2 A_n &= {\cal S}^A_n  \cr
  \left[ \partial_r^2 + {3 \over L_{\rm IR}} \partial_r \right]
   h_n &= {\cal S}^h_n  \cr
  \left[ \partial_r^2 + {1 \over L_{\rm IR}} \partial_r - 
    m_\Phi^2 \right] \Phi_n &= {\cal S}^\Phi_n  \cr
  \left[ \partial_r^2 + {3 \over L_{\rm IR}} \partial_r - 
    m_{\rm IR}^2 \right] \psi_n &= {\cal S}^\psi_n \,,
 }
where ${\cal S}^X_n$, for $X=A$, $h$, $\Phi$, or $\psi$ is a polynomial in the coefficient functions $A_k$, $h_k$, $\Phi_k$, and $\psi_k$, and their derivatives, for $k < n$.  ${\cal S}^X_1 = 0$ for $X=A$, $h$, $\Phi$, and $\psi$.  We choose $\Phi_1$ and $\psi_1$ as in \eno{AllowedSolns}, and we set $A_1=h_1=0$.  For $n>1$, the equations \eno{nthConditions} can be solved iteratively using a method of Green's functions:
 \eqn{GFsolve}{
  A_n(r) &= \int_{-\infty}^r d\tilde{r}
     \int_{-\infty}^{\tilde{r}} dr_* \, {\cal S}^A_n(r_*)
    = \int_{-\infty}^r dr_* \, (r-r_*) {\cal S}^A_n(r_*)  \cr
  h_n(r) &= \int_{-\infty}^r dr_* \, L_{\rm IR} 
   {1-e^{3(r_*-r)/L_{\rm IR}} 
    \over 3} {\cal S}^h_n(r_*)  \cr
  \Phi_n(r) &= \int_{-\infty}^r dr_* \, L_{\rm IR}
    {e^{(\Delta_\Phi-2)(r-r_*)/L_{\rm IR}} - 
      e^{-(\Delta_\Phi-1) (r-r_*)/L_{\rm IR}} \over 2\Delta_\Phi-3}
      {\cal S}^\Phi_n(r_*)  \cr
  \psi_n(r) &= \int_{-\infty}^r dr_* \, L_{\rm IR}
    {e^{(\Delta_{\rm IR}-3)(r-r_*)/L_{\rm IR}} - 
     e^{-\Delta_{\rm IR} (r-r_*)/L_{\rm IR}} \over
     2\Delta_{\rm IR} - 3} {\cal S}^\psi_n(r_*) \,.
 }
One can check that the solution \eno{GFsolve} satisfies the zero-energy constraint.  Heuristically, this is because the zero-energy constraint is trivially satisfied for the $AdS_4$ vacuum, and the perturbations \eno{GFsolve} are constructed so as to approach this limit as rapidly as possible as $r \to -\infty$.

Equation \eno{GFsolve} represents only one particular set of solutions to the equations \eno{nthConditions}.  All others can be obtained by adding solutions to the homogeneous equations.  For the purposes of studying irrelevant perturbations to the infrared $AdS_4$ vacuum, no such additions should be made.  To see this, first note that three of those solutions---$h_n = e^{-3r/L_{\rm IR}}$, $\Phi_n = e^{-(\Delta_\Phi-1) r/L_{\rm IR}}$, and $\psi_n = e^{-\Delta_{\rm IR} r/L_{\rm IR}}$---are disallowed because they diverge exponentially as $r \to -\infty$.  The two solutions $A_n = r$ and $h_n = 1$ can be added, but the zero-energy constraint imposes a relation between their coefficients, and when this constraint is satisfied, the effect of the addition is simply to change the normalization of $r$.  The solution $A_n=1$ need not be added because it can be offset by rescaling $t$ and $\vec{x}$.   This leaves only the solutions $\Phi_n = e^{(\Delta_\Phi-2) r/L_{\rm IR}}$ and $\psi_n = e^{(\Delta_{\rm IR}-3) r/L_{\rm IR}}$.  They are present for $n=1$ and need not be included at higher orders: Doing so would merely adjust the values of $a_\Phi$ and $a_\psi$.  

The parameters $L$, $m^2$, $u$, and $q$ that enter into the lagrangian \eno{Lagrangian} with the quartic potential \eno{VChoice} can be traded for $L_{\rm IR}$, $\psi_{\rm IR}$, $\Delta_\Phi$, and $\Delta_{\rm IR}$.  These four parameters, together with $a_\Phi$ and $a_\psi$, completely determine all the $X_n$, where as usual, $X$ denotes $A$, $h$, $\Phi$, or $\psi$.  For generic values of the parameters, the solutions take the form
 \eqn{Xsoln}{
  X_n = \sum_\alpha c^X_{n,\alpha} e^{-\gamma^X_{n,\alpha}
    r / L_{\rm IR}} \,,
 }
where $\alpha$ runs over some finite set, the $c_{n,\alpha}^X$'s are rational functions of the parameters (independent of $r$), and
 \eqn{gammaForm}{
  \gamma^X_{n,\alpha} = p^X_{n,\alpha} \Delta_\Phi 
    + s^X_{n,\alpha} \Delta_{\rm IR} + r^X_{n,\alpha} \,.
 }
The coefficients $p^X_{n,\alpha}$ and $s^X_{n,\alpha}$ are nonnegative integers (not both zero for a given value of $X$, $n$, and $\alpha$), and $r^X_{n,\alpha}$ are negative integers.

Clearly, the expansions \eno{FormalExpansions} are valid only when all the $\gamma^X_{n,\alpha}$ are positive.  The positivity constraints at level $n=1$ are $\Delta_\Phi > 2$ and $\Delta_{\rm IR} > 3$, which are trivial in the sense that they follow from the definitions \eno{DeltaIRdefs}.  At the quadratic level, $n=2$, one finds a tighter constraint from the $\gamma$ coefficients for $A$, $h$, and $\psi$: $\Delta_\Phi > 3$.  It is straightforward to check that the following two versions of the $n=2$ constraint are equivalent:
 \eqn{QuadraticConstraint}{
  \Delta_\Phi > 3 \qquad\iff\qquad q L_{\rm IR}
   \psi_{\rm IR} > 1 \,.
 }
If this constraint is violated, then there cannot be a domain wall interpolating between the $\psi=0$ and $\psi=\psi_{\rm IR}$ $AdS_4$ vacua.  No further tightening of constraints occurs at the next two orders, and we conjecture that there is no further tightening at any higher order, at least when the scalar potential is smooth.  Assuming this conjecture is correct, there is still a possibility of convergence problems in the infrared expansion, even though no individual term is badly behaved.  But numerical investigations suggest that a charged domain wall solution does exist, starting from the undeformed ultraviolet conformal theory, when the constraint \eno{QuadraticConstraint} is satisfied.

To recapitulate: The condition \eno{QuadraticConstraint}, in field theory terms, is the statement that the operator dual to $\Phi$ is irrelevant.  This is precisely the condition one expects in order for a flow to conformal invariance in the infrared to exist.  The series expansion machinery introduced in \eno{FormalExpansions}-\eno{GFsolve} confirms this expectation on the gravity side.  So we conclude that charged domain wall solutions with conformal invariance in the infrared probably exist when $\Delta_\Phi > 3$.

\section{Lifshitz-like scaling}
\label{LIFSHITZ}

The previous section, building upon results of \cite{Gubser:2008wz,Gubser:2009gp}, provides a candidate ground state of the Abelian Higgs model in $AdS_4$, provided there exists an extremum of the potential with $\psi \neq 0$, and provided the charge is not too small.  The candidate ground state is a domain wall interpolating between symmetry-preserving $AdS_4$ on the ultraviolet side and symmetry-breaking $AdS_4$ on the infrared side.  Its explicit form is given in \eno{AdSAnsatz}, with $h$ interpolating between two different constants in the ultraviolet and infrared, and with $\Phi$ vanishing in the infrared limit.  Slightly nonextremal generalizations of these domain walls would be approximately described as domain walls between the ultraviolet $AdS_4$ geometry and $AdS_4$-Schwarzschild in the infrared.

In this section, we propose another candidate ground state.  It is like the one just described in that it is a domain wall with symmetry-preserving $AdS_4$ in the ultraviolet.  But its infrared limit is a Lifshitz geometry similar to the ones constructed in \cite{Kachru:2008yh}.  In subsection~\ref{EMBED} we briefly review this construction and indicate how it can be formally embedded in a limit of the Abelian Higgs model.  In subsection~\ref{ONLY} we demonstrate that, besides $AdS_4$, $AdS_4$-Schwarzschild, and $AdS_4$-Reissner-Nordstrom, the Lifshitz geometry is the only solution to the equations of motion \eno{Aeom}-\eno{ZeroEnergy} that can have constant $\psi$.  In subsection~\ref{PERTURBATIONS}, we analyze the perturbations of Lifshitz backgrounds at linear order.  In subsection~\ref{QUADRATIC}, we discuss Lifshitz solutions based on the U-shaped quadratic potential: \eno{VChoice} with $m^2>0$ and $u=0$.\footnote{While this work was in progress, we were informed by M.~Roberts that he and G.~Horowitz have also studied the quadratic case.}  In subsection~\ref{QUARTIC} we discuss solutions with Lifshitz-like scaling based on the W-shaped quartic potential: \eno{VChoice} with $m^2<0$ and $u>0$.  

As in the case of emergent conformal symmetry discussed in the previous section, what we are doing here is constructing a geometry that {\it may} be the infrared side of a domain-wall ground state of the Abelian Higgs model in $AdS_4$.  To show that such domain walls really exist, the only approach we know of is numerics.  We give some examples in sections~\ref{QUADRATIC} and~\ref{QUARTIC}.

Altogether, our results on $AdS_4$-to-Lifshitz domain walls bear some resemblance to the work of \cite{Azeyanagi:2009pr}.  The main differences are that we do not attempt to embed our solutions into string theory, and that the coordinate in our solutions that scales anisotropically is not spatial but instead timelike.  (In \cite{Azeyanagi:2009pr}, a configuration was considered in which the scaling is anisotropic in the timelike direction.  However, this configuration involved a slightly unusual feature, namely a continuous density of extended fundamental strings.  So it is not wholly described in terms of supergravity, as our solutions are.)  Our domain walls are quite different from the one exhibited in \cite{Kachru:2008yh}, in that we have conformal invariance in the ultraviolet and Lifshitz behavior in the infrared, not the other way around.

\subsection{Embedding Lifshitz solutions in the Abelian Higgs model}\label{EMBED}

In \cite{Kachru:2008yh}, it was explained that four-dimensional gravity with a negative cosmological constant coupled to a two-form field strength $F_{(2)}$ and a three-form field strength $F_{(3)} = dB_{(2)}$, with a $B_{(2)} \wedge F_{(2)}$ interaction, admits solutions with Lifshitz-like symmetry.  The metric is
 \eqn{LifMet}{
  ds^2 = -\left( {r \over L_0} \right)^{2z} dt^2 + 
   {r^2 \over L_0^2} d\vec{x}^2 + L_0^2 {dr^2 \over r^2} \,,
 }
and the Lifshitz-like scaling symmetry is
 \eqn{ScalingSymmetry}{
  t \to \lambda^z t \qquad \vec{x} \to \lambda \vec{x} \qquad
    r \to {r \over \lambda} \,.
 }
The dynamical exponent $z$ is determined in terms of $L_0$ and the coupling multiplying the $B_{(2)} \wedge F_{(2)}$ term.  Note that in \eno{LifMet}, we have persisted in letting $r$ be a dimensionful variable.  To recover the form of the ansatz discussed in \cite{Kachru:2008yh}, one can use the dimensionless variables $t/L_0$, $\vec{x}/L_0$, and $r/L_0$ in place of $t$, $\vec{x}$, and $r$.

The $B_{(2)} \wedge F_{(2)}$ theory considered in \cite{Kachru:2008yh} is a limit of the Abelian Higgs model in which the modulus of $\psi$ is frozen at $\psi_{\rm IR}$.  Our main aim in this section is to explain how solutions of the form \eno{LifMet} arise in the Abelian Higgs model before any special limit is taken.  However, let us briefly detour to explain how to map the frozen modulus limit of the Abelian Higgs model into the $B_{(2)} \wedge F_{(2)}$ theory.  First, to define this limit, we consider a potential $V(\psi,\psi^*)$ that depends only on the modulus of $\psi$ and has a very sharp minimum at some finite value $\psi_0$ of $|\psi|$.  Restricting
 \eqn{psiExpress}{
  \psi = \psi_0 e^{i\theta} \,,
 }
one finds from \eno{Lagrangian} the action
 \eqn{BwedgeFstart}{
  S &= {1 \over 2\kappa^2} \int d^4 x \, \sqrt{-g} 
   {\cal L}
 }
where
 \eqn{Lform}{
  {\cal L} =  
    R - {1 \over 4} F_{\mu\nu}^2 - 
    \psi_0^2 (\partial_\mu \theta - q A_\mu)^2 - V_0
    - {q\sqrt{2} \psi_0 \over 4\sqrt{-g}}
     \epsilon^{\mu\nu\rho\sigma} B_{\mu\nu}
     \left( F_{\rho\sigma} - 2\partial_\rho A_\sigma
     \right) \,,
 }
and $V_0 = V(\psi_0,\psi_0)$.  In the last term of \eno{Lform} we have introduced a Lagrange multiplier field $B_{\mu\nu}$ that enforces $F_{(2)} = dA_{(1)}$ as a constraint.  When integrated against $\sqrt{-g}$, this term is topological in the sense that it does not involve the metric.  So it doesn't affect the Einstein equations.  The momentum conjugate to $\theta$ is
 \eqn{ConjugateTheta}{
  \Pi^\mu \equiv {\partial {\cal L} \over
    \partial (\partial_\mu \theta)} 
    = -2\psi_0^2 (\partial^\mu \theta - qA^\mu) \,,
 }
and it is conserved because $\theta$ enters into ${\cal L}$ only through its first derivatives.  A convenient way to express this conservation is
 \eqn{HodgePi}{
  \Pi^\mu = -{\sqrt{2} \psi_0 \over 3! \sqrt{-g}} 
    \epsilon^{\mu\nu\rho\sigma} F_{\nu\rho\sigma} \,,
 }
where $F_{(3)}$ is a closed three-form.  To obtain the equations of motion for the other degrees of freedom, one may use the Routhian construction:
 \eqn{Routhian}{
  {\cal R} \equiv {\cal L} - \Pi^\mu \partial_\mu \theta
    &= R - {1 \over 4} F_{\mu\nu}^2 + {1 \over 4\psi_0^2} \Pi_\mu^2 - 
      q \Pi^\mu A_\mu - V_0
    - {q\sqrt{2} \psi_0 \over 4\sqrt{-g}}
    \epsilon^{\mu\nu\rho\sigma} B_{\mu\nu}
     \left( F_{\rho\sigma} - 2\partial_\rho A_\sigma
     \right)  \cr
    &= R - {1 \over 4} F_{\mu\nu}^2 - {1 \over 12} 
      F_{\mu\nu\rho}^2 - V_0 - {q\sqrt{2} \psi_0 \over 4 \sqrt{-g}}
       \epsilon^{\mu\nu\rho\sigma} B_{\mu\nu} F_{\rho\sigma}  \cr
    &\qquad{} + {q\sqrt{2} \psi_0 \over 3! \sqrt{-g}} 
     \epsilon^{\mu\nu\rho\sigma} A_\mu (F_{\nu\rho\sigma} - 
       3 \partial_\nu B_{\rho\sigma}) + 
       \hbox{(total derivative)}\,.
 }
$A_\mu$ enters the final expression for the Routhian only as a lagrange multiplier enforcing the constraint $F_{(3)} = dB_{(2)}$.  We may omit the lagrange multiplier term if we elevate the constraint to a definition of $F_{(3)}$; then all equations of motion follow from the second line of \eno{Routhian}, and we have indeed recovered $B_{(2)} \wedge F_{(2)}$ theory.  The second line of \eno{Routhian} precisely matches (2.3) of \cite{Kachru:2008yh}, provided we set $\kappa=1$, $e=1$, $2\Lambda = V_0$, and $c = q\sqrt{2} \psi_0$.

\subsection{The uniqueness of Lifshitz solutions}\label{ONLY}

Having established that the $B_{(2)} \wedge F_{(2)}$ theory is equivalent to the frozen-modulus limit of the Abelian Higgs model, a natural follow-up question is whether the Lifshitz solutions \eno{LifMet} persist away from this limit.  We claim that they do, and that besides $AdS_4$, $AdS_4$-Schwarzschild, and $AdS_4$-Reissner-Nordstrom they are the only other solutions with translation invariance in time, spatial translation and rotation symmetry in the directions $x^1$ and $x^2$, and constant value for the scalar field $\psi$.  Without loss of generality, we can assume that this constant value is real and positive.  We will make this restriction from now on.

To establish our claim, we start with a general ansatz consistent with the symmetries mentioned:
 \eqn{BGansatz}{
  ds^2 &= -g(r)^2 dt^2 + {r^2 \over L_0^2} d\vec{x}^2 + 
    e^{2B(r)} {L_0^2 \over r^2} dr^2  \cr
  \Phi &= \Phi(r) \qquad\qquad \psi = \psi_0 = \hbox{(constant)} \,.
 }
The metric ansatz in \eno{BGansatz} is equivalent to the one in \eno{AdSAnsatz} after appropriate redefinitions of coordinates and fields.  Plugging the constant value $\psi_0$ into the scalar equation of motion results in the condition that 
 \eqn{VeffDef}{
  V_{\rm eff}(\psi,\psi^*) = V(\psi,\psi^*) - 
    {q^2 \Phi(r)^2 \psi\psi^* \over g(r)^2}
 }
is extremized at $\psi=\psi_0$.  There are two ways in which this can happen, for all $r$:
 \begin{enumerate}
  \item It could be that both terms of \eno{VeffDef} are separately extremized.
  \item It could be that neither term of \eno{VeffDef} is separately extremized, but instead that their first derivatives cancel at $\psi=\psi_0$.
 \end{enumerate}
Let us refer to solutions with constant $\psi$ as solutions of the first or second kind, according to which of the two possibilities just described is realized.  Because of the $U(1)$ symmetry, we can assume that $\psi_0$ is real and nonnegative.

For solutions of the first kind, we must have ${\partial V \over \partial\psi} = {\partial V \over \partial\psi^*} = 0$ at $\psi=\psi^*=\psi_0$,  and also that either $\Phi$ or $q\psi_0$ vanishes.  If $\Phi$ vanishes, then the solution \eno{BGansatz} can only be $AdS_4$ or $AdS_4$-Schwarzschild.  If $q\psi_0$ vanishes, then the possibilities are $AdS_4$ and $AdS_4$-Schwarzschild if $\Phi$ is constant, and $AdS_4$-Reissner-Nordstrom if it isn't.  Thus our claim comes down to demonstrating that solutions of the second type must exhibit Lifshitz-like scaling.

The condition that $V_{\rm eff}$ is extremized for all values of $r$ through non-trivial competition between the two terms is quite restrictive because it implies that 
 \eqn{gCoef}{
  g(r) = {1 \over \sqrt{2 - 2/\eta}} \Phi(r)
 }
for some constant $\eta$.  Plugging this equation into the Einstein equations, one can solve algebraically for $r\Phi''(r)$, $rB'(r)$, and $B(r)$ in terms of $\Phi(r)$, $r\Phi'(r)$, $V(\psi_0,\psi_0)$, $q\psi_0$, and $\eta$, with no additional dependence on $r$.  ($B''(r)$ doesn't enter to the Einstein equations because it is essentially a gauge degree of freedom.)  The resulting expressions are unenlightening, so we will not exhibit them explicitly.  Eliminating $r\Phi''(r)$, $rB'(r)$ and $B(r)$ from the Maxwell equation for $\Phi$, one obtains the relation
 \eqn{PhiRelate}{
  {V(\psi_0,\psi_0) \over q^2 \psi_0^2} = 
    {4\eta + 8z\eta + z^2 (-3+2\eta+\eta^2) \over
     \eta (-z^2+\eta+2z\eta)} \,,
 }
where 
 \eqn{PhiRoll}{
  z = {r \Phi'(r) \over \Phi(r)} \,.
 }
\eno{PhiRelate} shows that $z$ is a constant, so $\Phi$ has a power-law dependence, $\Phi \propto r^z$.  Plugging this dependence back into the Einstein equations leads to the constraint
 \eqn{zEtaConstraint}{
  (\eta-z)(z\eta+2\eta-z) = 0 \,.
 }
Assuming that the second factor vanishes leads to difficulties: The Maxwell equation for $\Phi$ then demands that $p_0 z (1+z) = 0$.  But if $z=0$ or $-1$ then $g_{rr}$ formally vanishes, while if $p_0=0$ then $g_{tt}$ formally vanishes.  So we may assume that $\eta=z$.  Then \eno{PhiRelate} becomes
 \eqn{FixVLif}{
  {V(\psi_0,\psi_0) \over q^2 \psi_0^2} = -{4 + z + z^2 \over z} \,,
 }
and the only additional constraints from the equations of motion are
 \eqn{FixVpLif}{
  {\partial V \over \partial\psi}(\psi_0,\psi_0) = 
    {\partial V \over \partial\psi^*}(\psi_0,\psi_0) = 
    2 {z-1 \over z} q^2 \psi_0 
 }
 \eqn{zRelation}{
  e^{2B} q^2 \psi_0^2 L_0^2 = z \,.
 }
The latter implies that $B$ is constant.  If one starts with a definite function $V(\psi,\psi^*)$ and a definite value of the charge $q$, then \eno{FixVLif} and \eno{FixVpLif} generically admit at most a discrete set of solutions for $z$ and $\psi_0$ (given that we assume that $\psi_0$ is real and nonnegative).  Then \eno{zRelation} can be regarded as determining the product $e^{2B} L_0^2$.  By themselves $e^{2B}$ and $L_0^2$ are not meaningful: A rescaling
 \eqn{FirstRescaling}{
  r \to \lambda_1 r \qquad L_0 \to \lambda_1 L_0 \qquad
    e^{2B} \to {1 \over \lambda_1^2} e^{2B}
 }
preserves the form of the ansatz \eno{BGansatz} and the product $e^{2B} L_0^2$.  We can use this scaling symmetry to set $B=0$.  A second scaling symmetry,
 \eqn{SecondRescaling}{
  r \to \lambda_2 r \qquad 
   \vec{x} \to {1 \over \lambda_2} \vec{x} \,,
 }
also preserves the form of the ansatz.  We know that $g(r) \propto r^z$, and use of the $\lambda_2$ symmetry allows us to dictate the constant of proportionality:
 \eqn{gChoice}{
  g = \left( {r \over L_0} \right)^z \qquad
  \Phi = \sqrt{2 - {2 \over z}} \left( {r \over L_0} \right)^z \,,
 }
where in the second equation we have used \eno{gCoef}.  Plugging \eno{gChoice} into \eno{BGansatz}, we recover the original ansatz \eno{LifMet} with Lifshitz-like scaling symmetry.  This completes our demonstration that geometries with Lifshitz-like scaling are the only possible solutions to the classical equations of motion following from \eno{Lagrangian}, other than $AdS_4$, $AdS_4$-Schwarzschild, and $AdS_4$-Reissner-Nordstrom, in which $\psi$ can be constant.

The Lifshitz solution we have described is a straightforward lift of the solution of \cite{Kachru:2008yh} to the Abelian Higgs model.  The relations \eno{FixVLif} and~\eno{zRelation} correspond precisely to relations obtained in the frozen-modulus limit, namely (2.11) (first line) and (2.7) of \cite{Kachru:2008yh}.  In order for $\Phi$ and $g$ both to be real, we must have $z \geq 1$ or else $z < 0$.  The latter possibility is ruled out by the relation \eno{zRelation}.  Having concluded that $z \geq 1$,\footnote{The inequality $z \geq 1$ was also obtained in \cite{Kachru:2008yh} for the $B_{(2)} \wedge F_{(2)}$ theory.  It can almost be obtained from a null-energy argument, as follows.  By calculation, $L_0^2 (-R^t_t + R^x_x) = (z+2)(z-1)$.  According to the Einstein equations, $-R^t_t + R^x_x = \kappa^2 (-T^t_t + T^x_x) = \kappa^2 T_{\mu\nu} \xi^\mu \xi^\nu$ where $\xi^\mu = \left({1 \over r^z},{1 \over r},0,0\right)$ is a null vector.  The null energy condition says that $T_{\mu\nu} \xi^\mu \xi^\nu \geq 0$ for any null vector $\xi^\mu$.  So $(z+2)(z-1) \geq 0$, implying either $z \leq -2$ or $z \geq 1$.  Nothing in this null-energy argument rules out $z \leq -2$, but \eno{zRelation} of course does.} we see from \eno{FixVpLif} that ${\partial V \over \partial\psi}(\psi_0,\psi_0) \geq 0$.  For the double-well quartic potential \eno{VChoice}, this is only true when $\psi_0 > \psi_{\rm IR}$, where $\psi_{\rm IR}$ is the positive real minimum of $V(\psi,\psi^*)$, as in \eno{psiIR}.  Finally, noting that $4+z+z^2 > 0$ when $z \geq 1$, we see from \eno{FixVLif} that $V(\psi_0,\psi_0) < 0$.  For the potential \eno{VChoice}, this implies $\psi_0 < \psi_*$ where $\psi_*$ is the unique positive root of the equation $V(\psi,\psi^*)=0$.

Although we did not use the Noether charge $Q$ defined in \eno{Noether} in our demonstration of uniqueness of Lifshitz backgrounds, it is straightforward, starting with the expression
 \eqn{Qagain}{
  Q = {1 \over g e^{B}} \left( {r \over L_0} \right)^3
    \left( 2gg' - {2g^2 \over r} - \Phi\Phi' \right) \,,
 }
to check that $Q$ does vanish.

\subsection{Perturbing a Lifshitz background}
\label{PERTURBATIONS}

In the previous subsection, we demonstrated that Lifshitz solutions to the equations of motion following from \eno{Lagrangian} exist precisely when we can simultaneously solve \eno{FixVLif} and \eno{FixVpLif}.  What we want to find out next is when such solutions can be matched onto an asymptotically $AdS_4$ geometry in order to describe a ground state of the asymptotically conformal holographic Abelian Higgs model.  The answer turns out to be a bit subtle for approximately the same reason that we encountered with emergent conformal symmetry: There may or may not be irrelevant perturbations to the Lifshitz background of a sort that allow it to participate in an $AdS_4$-to-Lifshitz domain wall.

To study perturbations, let's consider the ansatz \eno{BGansatz} again, but with $\psi$ allowed to be a function of $r$, and all functions expressed as perturbations of the Lifshitz solution \eno{LifMet}:
 \eqn{PerturbLif}{
  g &= \left( {r \over L_0} \right)^z + \lambda g_1 + \ldots  \cr
  B &= \lambda B_1 + \ldots  \cr
  \Phi &= \sqrt{2 - {2 \over z}} \left( {r \over L_0} \right)^z + 
    \lambda \Phi_1 + \ldots  \cr
  \psi &= \psi_0 + \lambda \psi_1 + \ldots \,,
 }
where $\lambda$ is a formal expansion parameters, and $g_1$, $B_1$, $\Phi_1$, and $\psi_1$ are functions only of $r$.  In the most general ansatz consistent with preservation of the translation symmetries in the $t$ and $\vec{x}$ direction and the rotation symmetry between $x^1$ and $x^2$, we would have to include also perturbations $\delta g_{xx}$ and $\delta g_{tr}$ to the metric and $\delta A_r$ to the gauge field.  Excluding these additional perturbations amounts to partially gauge-fixing.

The five functions $(g_1,B_1,\Phi_1,\psi_1,\psi_1^*)$ are subject to five second-order differential equations plus three first order constraints, obtained by linearizing the equations of motion in $\lambda$.  So there are seven linearly independent solutions.  Two of the seven perturbations are trivial:
 \begin{enumerate}
  \item[$1.$] $\psi_1 = -\psi_1^* = i$, corresponding to changing the background value of the scalar from $\psi_0$ to $e^{i\theta_0} \psi_0$, where $\theta_0$ is some constant phase.  (Recall we assume that $\psi_0$ is real and positive.)
  \item[$2.$] $g_1 = (r/L_0)^z$ and $\Phi = \sqrt{2-2/z} (r/L_0)^z$, corresponding to rescaling $t$ by a constant.
 \end{enumerate}
Both these two pure gauge modes, and the other five perturbations, can be put into the general form
 \eqn{PowerScaling}{
  g_1 = c_g r^{\beta_g} \qquad
   B_1 = c_B r^{\beta_B} \qquad
   \Phi_1 = c_\Phi r^{\beta_\Phi} \qquad
   \delta\psi = c_\psi r^{\beta_\psi} \qquad
   \delta\psi^* = c_{\psi^*} r^{\beta_{\psi^*}} \,,
 }
and one always finds the following relations among the exponents:
 \eqn{betaRelate}{
  \beta_g = \beta_B + z = 
    \beta_\psi + z = \beta_{\psi^*} + z = \beta_\Phi \,.
 }
In order for a perturbation to be ``irrelevant,'' all the ${\cal O}(\lambda)$ corrections should become small compared to the leading order solution, except for $B_1$, which should become small compared to $1$.  This happens precisely if $\Re\beta_\psi > 0$.  If instead $\Re\beta_\psi < 0$, then the perturbation is ``relevant'' in the sense of becoming larger as one passes toward the infrared.

The remaining five perturbations fall into two classes (c.f.~the analysis of \cite{Bertoldi:2009vn}):
 \begin{enumerate}
  \item[$3.$] There is one perturbation with $\beta_\psi = -2-z$, which we will term the ``universal mode.''  One can show that
 \eqn{cUniversal}{
  c_{tt} &= -L_0^{-z} \sqrt{z(z-1) \over 2}
    {2(z^2+9z+2) + \psi_0^2 m_0^2 L_0^2 (z-2) \over
     6 z (z-1) + \psi_0^2 m_0^2 L_0^2 (z^2+2)} c_\Phi  \cr
  c_{rr} &= -L_0^{2+z} \sqrt{z(z-1) \over 2}
    (z+2) {2(z+3) + \psi_0^2 m_0^2 L_0^2 \over
     6 z (z-1) + \psi_0^2 m_0^2 L_0^2 (z^2+2)} c_\Phi  \cr
  c_\psi &= c_{\psi^*} = 
    L_0^z \sqrt{z(z-1) \over 2} {2(z+1)(z+2) \psi_0 \over
     6 z (z-1) + \psi_0^2 m_0^2 L_0^2 (z^2+2)} c_\Phi \,,
 }
where we have defined
 \eqn{mZeroDef}{
  m_0^2 \equiv 
   {\partial^2 V \over \partial\psi \partial\psi^*}(\psi_0,\psi_0) + 
   {\partial^2 V \over \partial\psi^{*2}}(\psi_0,\psi_0) \,.
 }
We describe this mode as universal because it is present even in the frozen modulus limit where $m_0^2 \to \infty$.  In that limit, one can see from \eno{cUniversal} that if $c_{tt}$, $c_{rr}$, and $c_\Phi$ are held finite, then $c_\psi \to 0$, indicating that the scalar stays pinned at its background value.

Although the universal mode is a solution of the linearized equations of motion, it is not a solution of the linearization of the extremality condition $Q=0$.  Its interpretation seems to follow fairly clearly: This mode is related to making Lifshitz backgrounds nonextremal.  This is confirmed in the $B_{(2)} \wedge F_{(2)}$ theory by the calculations of \cite{Bertoldi:2009dt}: See in particular the ultraviolet asymptotics of the non-extremal backgrounds constructed there.
 \item[$4.$] There are four perturbations that we will term ``non-universal'' because their characteristics depend on details of the potential.  Each one is based on one of the following values of $\beta_\psi$:
 \eqn{BetapsiValues}{
  \beta_\psi = \beta_\psi(s_1,s_2) \equiv
   -{z+2 \over 2} + {s_1 \over \psi_0}
    \sqrt{D_1 + s_2 \sqrt{D_2}} \,,
 }
where
 \eqn{DeltaDefs}{
  D_1 &= -z+1 + \left( {5z^2 \over 4} - 2z + 3 + 
    {m_0^2 L_0^2 \over 2} \right) \psi_0^2 \cr
  D_2 &= \left[\left(z^2-3z+2-{m_0^2 L_0^2 \over 2}\right)\psi_0^2+z-1\right]^2-8\left(z^2-1\right)\psi_0^2
 }
and $s_1$ and $s_2$ are independently chosen to be either $+1$ or $-1$.  The values \eno{BetapsiValues} come out of insisting that when $c_\psi = c_{\psi^*}$ and $\beta_\psi$ is neither $-2-z$ nor $0$, the determinant of the matrix that constrains $c_{tt}$, $c_{rr}$, $c_\Phi$, and $c_\psi$ must vanish.  The closed-form expressions for these coefficients are long and unenlightening.  It is easily seen that the real parts of both $\beta_{\psi}(-1, 1)$ and $\beta_{\psi}(-1, -1)$ are always negative. Therefore, these exponents are always associated with relevant perturbations, and they never participate in a flow toward an infrared Lifshitz fixed point. On the other hand, the $\beta_{\psi}(1,\pm 1)$ may be associated with relevant or irrelevant perturbations, as we will see in the next section.
 \end{enumerate}

\subsection{The positive mass quadratic potential}
\label{QUADRATIC}

Inspecting \eno{FixVpLif} and recalling that we have to have $z \geq 1$, we see that $V$ has to slope upward in the direction of increasing magnitude of $\psi$ in order for there to be a Lifshitz fixed point. The simplest nontrivial potential for $\psi$ that satisfies this upward slope condition is the positive mass quadratic potential:
 \eqn{Vquadratic}{
  V(\psi,\psi^*) = -{6 \over L^2} + m^2 \psi\psi^*
 }
with $m^2 > 0$. Since the limiting case $z=1$ corresponds to $AdS_4$, we restrict ourselves to $z>1$. Solving \eno{FixVLif}, \eno{FixVpLif}, and \eno{zRelation} simultaneously with $B=0$, one can readily show that
 \eqn{zPsiL0}{
  q^2 = {zm^2\over 2(z-1)} \qquad
  \psi_0 = {2\sqrt{3} \over mL} \sqrt{{z-1\over (z+1)(z+2)}} \qquad
  L_0=L\sqrt{{(z+1)(z+2)\over 6}}\,\,.
 }
So every ordered pair $(z>1, m^2L^2>0)$ corresponds to a unique Lifshitz solution, and the ordered pairs $(z>1, m^2L^2>0)$ span the space of Lifshitz solutions admitted by positive mass quadratic potentials. But every choice of $(z>1, m^2L^2>0)$ doesn't necessarily permit a ``superconducting'' flow from a conformal fixed point in the ultraviolet to a Lifshitz fixed point in the infrared.  Ultimately, it appears to require numerical work to determine precisely when such a flow exists.  However, two complementary lines of thought provide important partial insight into when such flows should exist:
\begin{itemize}
\item The boundary geometry of extremal RNAdS in four dimensions is $AdS_4$, but the near-horizon geometry is $AdS_2\times \textbf{R}^2$. Though the complex scalar $\psi$ satisfies the Breitenlohner-Freedman (BF) bound \cite{Breitenlohner:1982bm,Breitenlohner:1982jf} $m^2L^2>-9/4$ at the boundary, it may not satisfy the analogous bound in $AdS_2$ near the horizon. If it doesn't, there is an instability, which suggests that the complex scalar $\psi$ assumes a nontrivial profile and spontaneously breaks the Abelian gauge symmetry. Similar arguments can be found in earlier works, including~\cite{Gubser:2005ih,Hartnoll:2008kx,Gubser:2008pf,Denef:2009tp,Faulkner:2009wj}. The derivation below closely follows the development in~\cite{Gubser:2008pf}. Using the metric convention \eno{AdSAnsatz}, the RNAdS solution is
\eqn{RNAdSMetric}{
A &={r\over L}\qquad h=1-\epsilon L \kappa^2 e^{-3r/L}+{\rho^2\kappa^4\over 4} e^{-4r/L} \cr
\Phi &=\rho\kappa^2(e^{-r_H/L}-e^{-r/L})\qquad \psi=0\,,
}
where the horizon $r=r_H$ occurs where $h=0$. We are free to set $r_H=0$. At extremality, $h$ has a double zero at $r=0$, and
\eqn{ExtremalRNAdS}{
\rho={2\sqrt{3}\over\kappa^2}\qquad\epsilon={4\over\kappa^2 L}\,.
}
It is now straightforward to show that near $r=0$, the extremal metric takes the form
\eqn{HorExtremalRNAdS}{
ds^2=\underbrace{-{r^2\over (L/\sqrt{6})^2}dt^2+{(L/\sqrt{6})^2\over r^2}dr^2}_{AdS_2}+\underbrace{dx_1^2+dx_2^2}_{\textbf{R}^2}\,.
}
The curvature radius $L_{AdS_2}$ of near-horizon $AdS_2$ is thus
\eqn{LAdS2}{
L_{AdS_2}=L/\sqrt{6} \,.
}
The BF bound in near-horizon $AdS_2$ is violated when
\eqn{BFAdS2}{
m_{AdS_2}^2 L_{AdS_2}^2 < -{1\over 4}\,.
}
Above, $m_{AdS_2}^2$ is the limit $r\to 0$ of the effective mass squared $m_{\rm eff}^2$ of $\psi$, which was defined in~\cite{Gubser:2008px} as
\eqn{EffMass}{
m_{\rm eff}^2=m^2+g^{tt}q^2\Phi^2\,.
}
Plugging the RNAdS metric \eno{RNAdSMetric} into this equation and taking the limit $r\to 0$, we find that
\eqn{mAdS2}{
m_{AdS_2}^2=m^2-2q^2.
}
So the BF bound in near-horizon $AdS_2$ suggests that extremal RNAdS is unstable when
\eqn{FinalBFCondition}{
m^2L^2-2q^2L^2<-{3\over 2}\,.
}
(It is interesting to note that the first relation in \eno{zPsiL0} implies that Lifshitz solutions for the positive mass potential exist only when $m^2L^2-2q^2L^2<0$.)  \eno{FinalBFCondition} translates to an inequality relating $m^2L^2$ and $z$ when we plug in the expression for $q^2$ from \eno{zPsiL0}:
\eqn{m2zBFCondition}{
m^2L^2>{3\over 2}(z-1) \,.
}
When this inequality is obeyed, we have an {\it a priori} reason to expect there are symmetry-breaking solutions with nonzero $\psi$: the instability of extremal RNAdS. When this inequality is not obeyed, there is no known instability in extremal RNAdS, and the existence of symmetry-breaking solutions is less likely. Indeed, we have been unable to numerically construct symmetry-breaking solutions that violate \eno{m2zBFCondition}.

\item There are two non-universal perturbations of a given Lifshitz fixed point that can be irrelevant. They are associated with the powers $\beta_{\psi}(1,\pm 1)$.  At least one perturbation must be irrelevant in order for the Lifshitz point to participate in a flow from an ultraviolet conformal field theory to infrared Lifshitz behavior, simply because the approach to Lifshitz behavior is described by some irrelevant perturbation.  Usually we are interested in flows to infrared Lifshitz behavior that arise spontaneously from a conformal fixed point: that is, we prescribe that there is no explicit symmetry breaking in the ultraviolet.  In order to impose such a constraint, a generic expectation is that one must have not one but two irrelevant perturbations in the infrared, so that one parameter (besides an overall energy scale) can be tuned in the infrared to accommodate the constraint in the ultraviolet.  This generic expectation might fail at a co-dimension one locus in the space of allowed $(z,m^2L^2)$.

An interesting possibility is that $\beta_\psi(1,1)$ and $\beta_\psi(1,-1)$ could be complex.  If they are, then they are complex conjugates of one another, and in order to have an $AdS_4$-to-Lifshitz flow, their real parts must be positive.  Keeping in mind $m_0=m$ for a quadratic potential, we can plug the expressions for $\psi_0$ and $L_0$ from \eno{zPsiL0} into \eno{BetapsiValues} to obtain
\eqn{BetasQuad}{
 \beta_{\psi}(1,\pm 1)=\beta_{\psi}^{\rm quad}(1,\pm 1)\equiv-{z+2\over 2}+\sqrt{d_1\pm\sqrt{d_2}}\,,
 }
 where
 \eqn{Betads}{
  d_1 &= {5\over 4}z^2-2z+3\cr
  d_2 &= 12(z-1)^2 (z-2)^2-8m^2 L^2 (z+1)^2 (z+2) \,.
 }
 The quantity $d_1$ is always positive and greater than $\sqrt{d_2}$ for positive $d_2$. It follows that the $\beta_{\psi}^{\rm quad}(1,\pm 1)$ are only complex when $d_2$ is negative, and that there is a transition from real $\beta_{\psi}^{\rm quad}(1,\pm 1)$ to complex $\beta_{\psi}^{\rm quad}(1,\pm 1)$ where $d_2$ vanishes.
More specifically, the $\beta_{\psi}^{\rm quad}(1,\pm 1)$ are only real when
\eqn{QuadConditionRC}{
m^2 L^2<{3\over 2}{(z-2)^2(z-1)^2\over (z+1)^2 (z+2)} \,.
}
One can also easily show that $\Re \beta_\psi^{\rm quad}(1,\pm 1) > 0$ precisely when
\eqn{QuadConditionRP}{
m^2 L^2>{3 \over 2} {(2-z)(z-1)(7z+2)\over (z+1)^2(z+2)} \,.
}
If the $\beta_\psi^{\rm quad}(1,\pm 1)$ are real, then to determine where in parameter space there are irrelevant deformations of Lifshitz solutions, we should ask when $\beta_\psi^{\rm quad}(1,\pm 1)$ vanishes.  It is easily checked that
\begin{itemize}
\item[$\circ$] $\beta_{\psi}^{\rm quad}(1, 1)$ only vanishes when $z\le 2$ and $m^2L^2\to 0$.
\item[$\circ$] $\beta_{\psi}^{\rm quad}(1, -1)$ only vanishes when $z\ge 2$ and $m^2L^2\to 0$.
\end{itemize}
So a critical point occurs at $(z, m^2L^2)=(2,0)$, where both $\beta_{\psi}^{\rm quad}(1, 1)$  and $\beta_{\psi}^{\rm quad}(1, -1)$ are zero. This critical point coincides with a minimum of the RHS of the inequality \eno{QuadConditionRC}, and it is also where the RHS of the inequality \eno{QuadConditionRP} crosses the $z$-axis in $z$-$m^2L^2$ space. 
\end{itemize}

\begin{figure}
  \centerline{\includegraphics[width=7in]{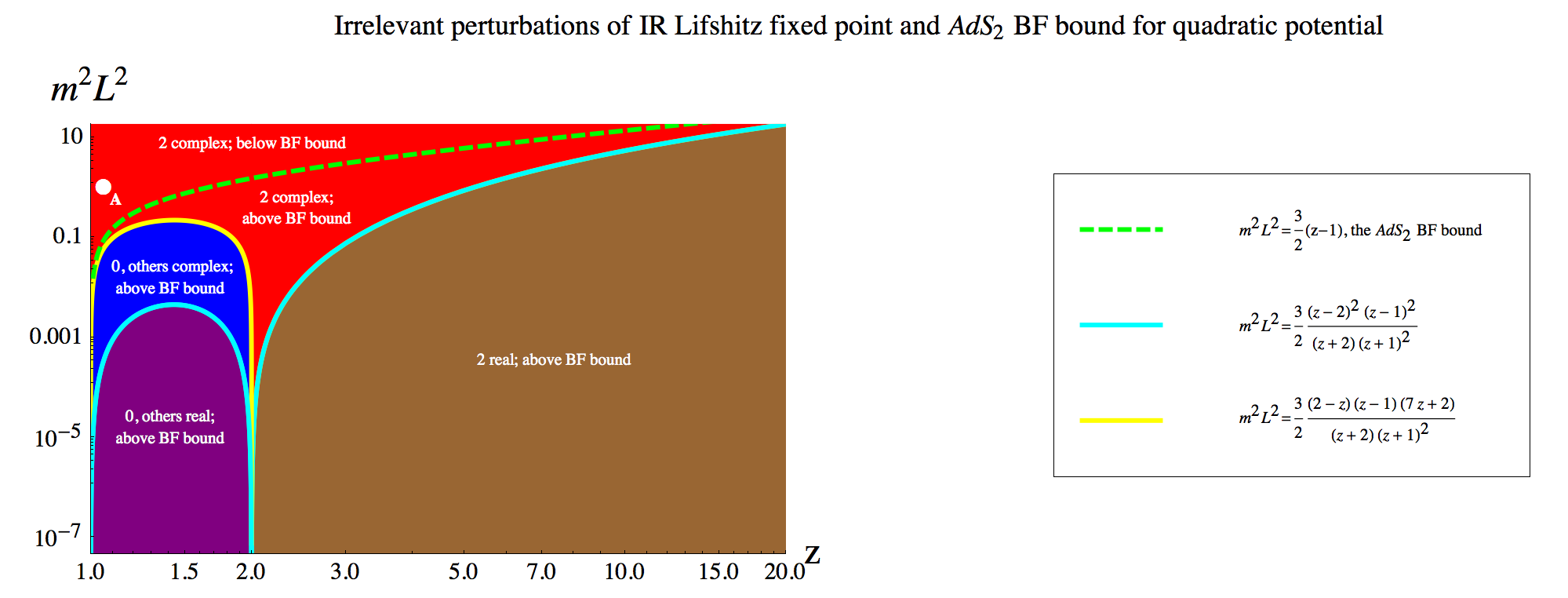}}
  \caption{(COLOR ONLINE) The number of irrelevant perturbations to the Lifshitz solution for a quadratic potential, as a function of $z>1$ and $m^2 L^2 > 0$. The two powers $\beta_\psi(1,1)$ and $\beta_\psi(1,-1)$ that characterize infrared  perturbations away from this solution fall into one of the four categories described in the text. The four categories meet at the point $(z, m^2L^2)=(2, 0)$. Point $A$ corresponds to an example flow discussed in the text and displayed in figure~\ref{ColdBH}.}\label{QuadPhase}
 \end{figure}

Figure~\ref{QuadPhase} ties together all the features of the above discussion by plotting the $AdS_2$ BF bound presented in \eno{m2zBFCondition} and dividing $z$-$m^2L^2$ space into four categories:
\begin{enumerate}
\item (purple) Both $\beta_{\psi}^{\rm quad}(1, 1)$ and $\beta_{\psi}^{\rm quad}(1, -1)$ are real and negative. They correspond to relevant perturbations.
\item (brown) Both $\beta_{\psi}^{\rm quad}(1, 1)$ and $\beta_{\psi}^{\rm quad}(1, -1)$ are real and positive. They correspond to irrelevant perturbations.
\item (red) $\beta_{\psi}^{\rm quad}(1, 1)=\beta_{\psi}^{\rm quad}(1, -1)^*$ is complex with a positive real part. The $\beta_{\psi}^{\rm quad}(1, \pm 1)$ correspond to irrelevant perturbations. A flow to a conformal fixed point in the UV would exhibit damped oscillations in the IR.
\item (blue) $\beta_{\psi}^{\rm quad}(1, 1)=\beta_{\psi}^{\rm quad}(1, -1)^*$ is complex with a negative real part. The $\beta_{\psi}^{\rm quad}(1, \pm 1)$ correspond to relevant perturbations.
\end{enumerate}

Evidently, only complex $\beta_{\psi}(1,\pm 1)$ associated with irrelevant perturbations obey the inequality \eno{m2zBFCondition}.  Therefore, the positive mass quadratic potential probably admits $AdS_4$-to-Lifshitz flows only in cases where the approach to the Lifshitz point is oscillatory.

 \begin{figure}
  \centerline{\includegraphics[width=6in]{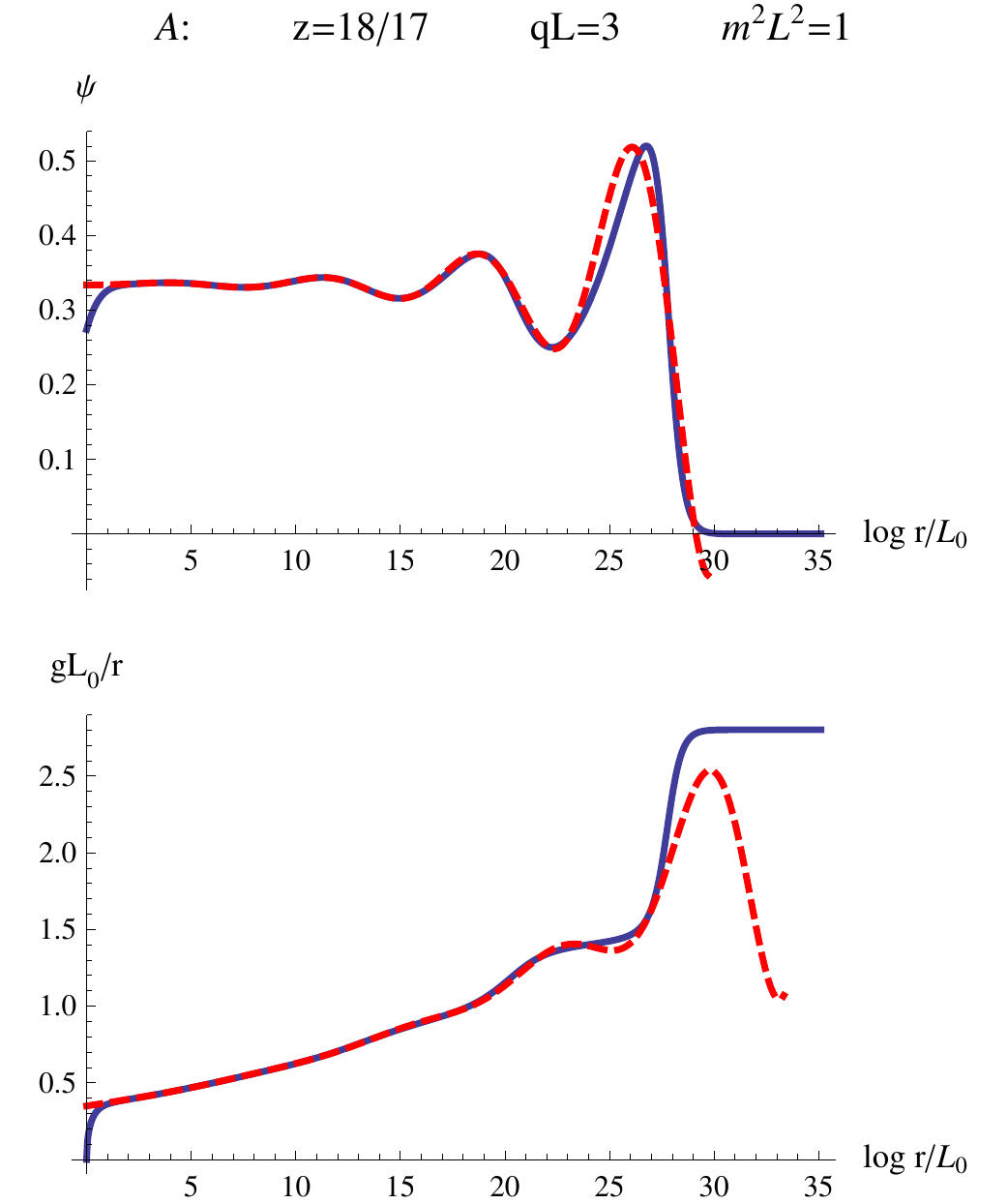}}
  \caption{(COLOR ONLINE) The blue curves are $\psi$ and $gL_0/r$ for a very cold superconducting black hole based on the positive mass quadratic potential with $qL = 3$ and $m^2 L^2 = 1$.  The temperature of this black hole is $T/\mu\approx 2.356\times 10^{-14}$, twelve orders of magnitude lower than the highest temperature at which the Abelian gauge symmetry is broken by $\psi$, $T_c/\mu\approx 0.0864$. The dotted red curves represent near-horizon fits to zero-temperature ansatzes that describe perturbations away from an infrared Lifshitz fixed points.}\label{ColdBH}
 \end{figure}

As we have already remarked, the considerations going into figure~\ref{QuadPhase} provide only partial insight into when $AdS_4$-to-Lifshitz flows exist, and the only way we know of definitely establishing existence is to construct such flows numerically.  In previous works we have pursued two different numerical strategies:
 \begin{enumerate}
  \item One can construct the zero-temperature solution directly, as in \cite{Gubser:2008wz}, provided one has analytic control over the infrared asymptotics.  This ``direct'' approach to constructing candidate ground states of the holographic Abelian Higgs model has the advantage of speed and simplicity.
  \item One can find the hottest $AdS_4$-Reissner-Nordstrom solution with a static solution to the linearized equation for $\psi$ and then follow the branch of solutions with $\psi \neq 0$ down toward extremality, as in \cite{Gubser:2008pf}.  Although more laborious than the direct approach, this ``cooling'' approach has the advantage that one knows how the symmetry-breaking ground state connects to the phase with unbroken symmetry.
 \end{enumerate}
We pursued the cooling approach to generate a very cold black hole solution to the theory with $qL=3$ and $m^2 L^2 = 1$, which corresponds to point $A$ in figure~\ref{QuadPhase}. In figure~\ref{ColdBH} we compare the numerically obtained $\psi(r)$ and $g(r)$ with fits to the expected zero-temperature behavior.  The blue curves represent the low-temperature solution for $qL=3$ and $m^2L^2=1$, which corresponds to point $A$ in figure~\ref{QuadPhase}.  In the corresponding zero-temperature solution for $\psi$, the irrelevant perturbations of the Lifshitz fixed point are characterized by the powers $\beta_{\psi}(1,1)=\beta_{\psi}(1,-1)^*\approx 0.204+0.848i$. The dotted red line in the plot of $\psi$ is a fit of the zero-temperature ansatz
\eqn{PsiAnsatz}{
\psi(r)=\psi_0+c_{\psi}r^{\beta_\psi(1,1)}+c_{\psi}^{*}r^{\beta_\psi(1,-1)}
}
to the behavior of the low-temperature solution close, but not too close, to the horizon.  (In practice, this meant for $\log r/L_0$ approximately between $1$ and $8$.)
Above, $\psi_0=\sqrt{51/455}$ and $z=18/17$, as can be determined from \eno{zPsiL0}. The fit parameters are the real and imaginary parts of $c_{\psi}$.  The dotted red line in the plot of $gr/L_0$ is based on the infrared asymptotics with the same value of $c_\psi$.  An overall scale factor in $g$ can be adjusted as a consequence of a symmetry of the equations of motion:
\eqn{gfitsymm}{
g\to cg \qquad \Phi\to c\Phi \,.
}
We fix this scale factor using a fit to the low-temperature solution. The agreement between low-temperature numerics and the analytic zero-temperature asymptotics is evidently excellent, except extremely close to the horizon (i.e.~for $0 < \log r/L_0 \lsim 1.5$), where finite-temperature effects become important, and far from the horizon (i.e.~for $\log r/L_0 \gsim 25$), where the roll-over from Lifshitz behavior to the ultraviolet conformal behavior occurs.

\subsection{The W-shaped quartic potential}
\label{QUARTIC}

We argued in section~\ref{CONFORMAL} that domain wall solutions with conformal invariance in both the ultraviolet and the infrared probably exist provided $q L_{\rm IR} \psi_{\rm IR} > 1$, which is equivalent to $\Delta_\Phi > 3$.  Suppose we hold the potential fixed (so that in particular $L_{\rm IR}$ and $\psi_{\rm IR}$ are fixed) and lower $q$ below the value permitted by this inequality.  What happens to the domain wall solutions?  We expect that they still exist, but have Lifshitz-like symmetry in the infrared instead of emergent conformal symmetry.  A heuristic reason to think this is the right idea is that when $\Delta_\Phi = 3$, the second order corrections to the solution include a constant shift of $\psi$ away from $\psi_{\rm IR}$.  So it seems sensible that the system would find a different solution with constant $\psi$.  As we saw in subsection~\ref{ONLY}, Lifshitz scaling is the only possibility.  It can be further checked that the second order shift of $\psi$ away from $\psi_{\rm IR}$ is positive when $\Delta_\Phi = 3$, which makes sense since Lifshitz solutions exist only in the region where the potential slopes upward.  

\begin{figure}
  \centerline{\includegraphics[width=6in]{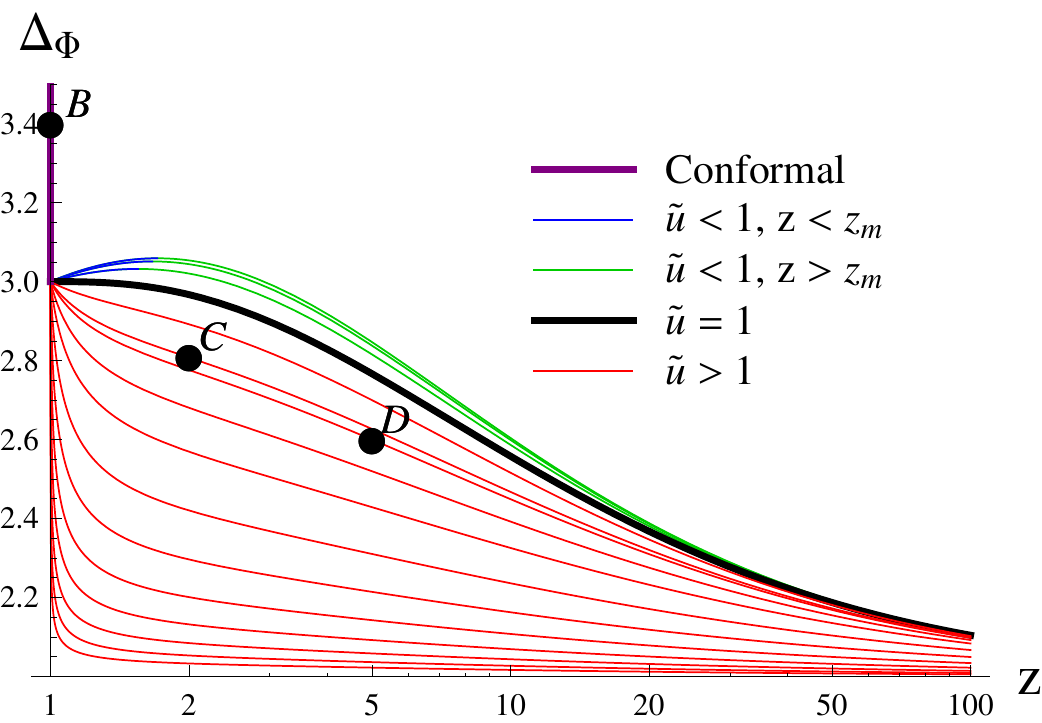}}
  \caption{(COLOR ONLINE) The behavior of $\Delta_{\Phi}$ as a function of $z$ for the quartic potential \eno{VChoice}.  The vertical purple line reminds us that for $\Delta_{\Phi} > 3$, domain walls with emergent conformal symmetry are allowed.  The various curves show how $\Delta_{\Phi}$ behaves as a function of $z$ in backgrounds with Lifshitz scaling.  Each curve corresponds to a definite value of the rescaled quartic coupling $\tilde{u}$.  For $\tilde{u} < 1$, $z_m$ is the value of $z$ where $\Delta_{\Phi}$ is maximized.}\label{DoublePhase}
 \end{figure}

\begin{figure}
  \centerline{\includegraphics[width=6.9in]{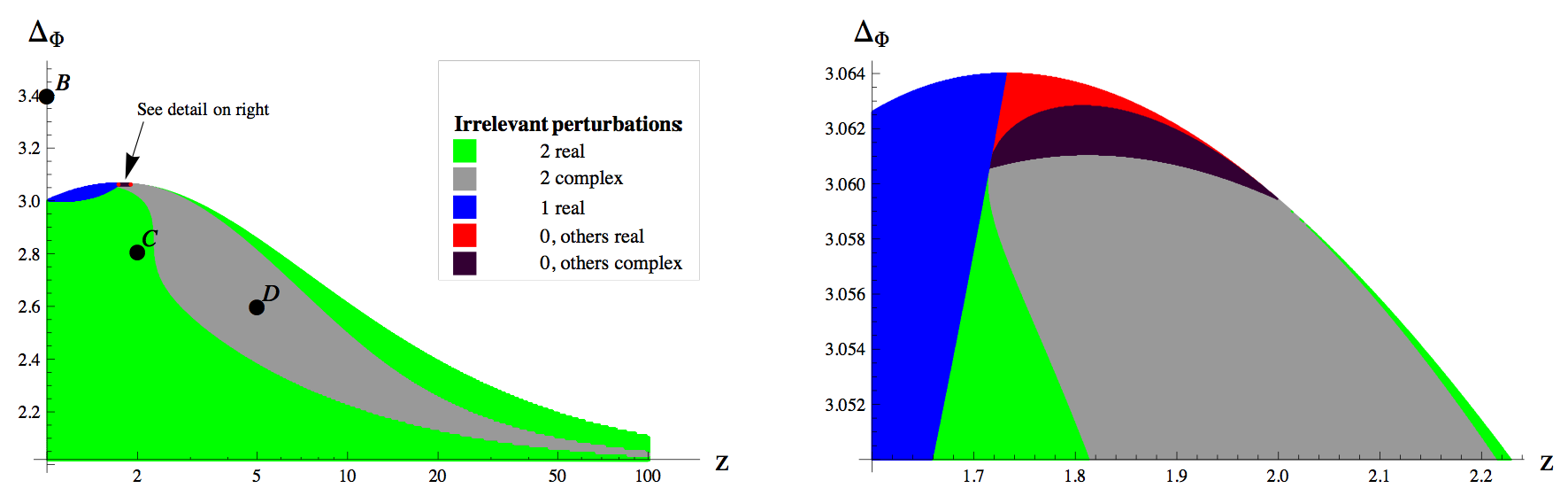}}
  \caption{(COLOR ONLINE) When there is a Lifshitz solution at a given point $(z, \Delta_{\Phi})$, the two powers $\beta_{\psi}(1,1)$ and $\beta_{\psi}(1,-1)$ that characterize perturbations away from this solution in the infrared fall into one of the five categories described in the text and summarized briefly in the legend. In the plot above, we have taken $m^2=-2$ and $L=1$.  The detail on the right shows that the five categories meet at the point $(z, \Delta_{\Phi})\approx (1.715, 3.061)$.}\label{PhaseDiagram}
 \end{figure}

\begin{figure}
  \centerline{\includegraphics[width=6in]{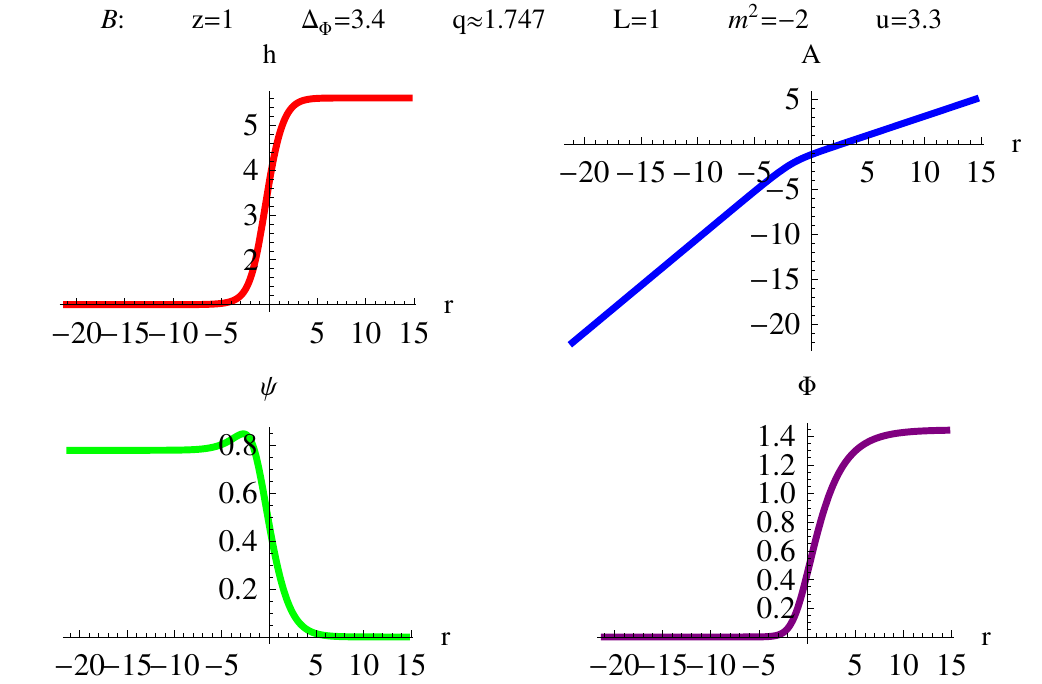}}
  \caption{(COLOR ONLINE) A flow between two conformal fixed points.}\label{BFlow}
 \end{figure}

\begin{figure}
  \centerline{\includegraphics[width=6in]{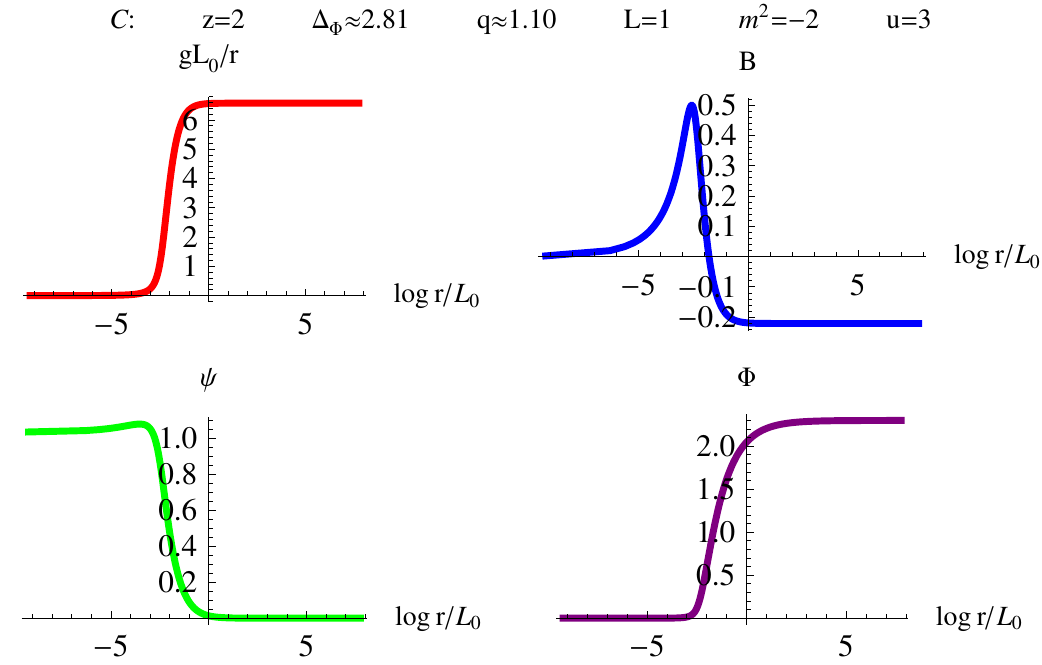}}
  \caption{(COLOR ONLINE) A flow between a Lifshitz fixed point in the infrared and a conformal fixed point in the ultraviolet for real $\beta_{\psi}(1,1)$ and $\beta_{\psi}(1,-1)$. There are no oscillations in the infrared.}\label{CFlow}
 \end{figure}

\begin{figure}
  \centerline{\includegraphics[width=6in]{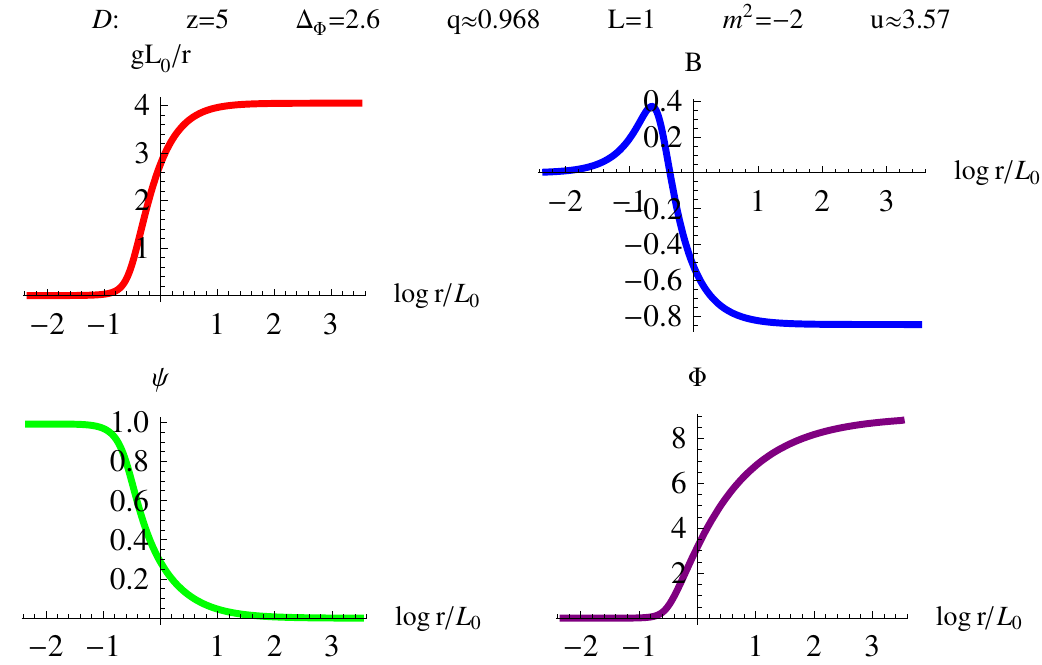}}
  \caption{(COLOR ONLINE) A flow between a Lifshitz fixed point in the infrared and a conformal fixed point in the ultraviolet for complex $\beta_{\psi}(1,1)$ and $\beta_{\psi}(1,-1)$. The damped oscillations in the deep infrared are imperceptible.}\label{DFlow}
 \end{figure}

The rest of this section is structured as follows.  First we give an analysis, for the W-shaped quartic potential, of when Lifshitz solutions exist.  The results are summarized in figure~\ref{DoublePhase}.  Next we present a (mostly) analytical study of whether there are irrelevant perturbations to the Lifshitz solutions.  The outcome of this study is shown in figure~\ref{PhaseDiagram}.  Finally, in figures~\ref{BFlow}-\ref{DFlow} we provide one explicit example of an $AdS_4$-to$AdS_4$ flow and two examples of $AdS_4$-to-Lifshitz flows.

Observe that with the help of \eno{LIRdef}, \eno{FixVLif} and \eno{FixVpLif} can be brought into the form
 \eqn{Justzp}{
  -{\psi_0 \over V(\psi_0,\psi_0)} {\partial V \over \partial\psi^*}
     (\psi_0,\psi_0) = 2 {z-1 \over z^2 + z + 4}
 }
 \eqn{qpL}{
  q^2 \psi_{\rm IR}^2 L_{\rm IR}^2 = 
    {6z \over 4+z+z^2} {V(\psi_0,\psi_0) \over V(\psi_{\rm IR},\psi_{\rm IR})}
     {\psi_{\rm IR}^2 \over \psi_0^2} \,.
 }
Specializing to the quartic potential \eno{VChoice} and defining
 \eqn{yDef}{
  y \equiv {\psi_0 \over \psi_{\rm IR}} \qquad\qquad
  \tilde{u} = {6u \over m^4 L^2} \,,
 }
we find that \eno{Justzp} and \eno{qpL} can be rewritten as
 \eqn{JustzpAgain}{
  {4y^2(y^2-1) \over 2 \tilde{u} + 2y^2 - y^4} = 
    4 {z-1 \over 4 + z + z^2}
 }
 \eqn{qpLagain}{
  q^2 \psi_{\rm IR}^2 L_{\rm IR}^2 = 
     {2\tilde{u} + 2y^2 - y^4 \over y^2 (1+2\tilde{u})}
     {6z \over 4 + z + z^2} \,.
 } 
If we also define $y_* = \psi_*/\psi_{\rm IR}$, then it is straightforward to check that for $\tilde{u} > 0$ and $z > 1$, there is a unique solution $y$ to \eno{JustzpAgain} with $1 < y < y_*$.  This is the allowed range of $y$ because it corresponds to values of $\psi$ between the minimum of $V(\psi,\psi^*)$ at $\psi_{\rm IR}$ and its zero at $\psi_*$.  Plugging this solution $y$ of \eno{JustzpAgain} into \eno{qpLagain}, one obtains a unique value for $q^2 \psi_{\rm IR}^2 L_{\rm IR}^2$, and hence a definite prediction for $\Delta_\Phi$, based on \eno{mIRdefs} and~\eno{DeltaIRdefs}, at the symmetry-breaking conformal fixed point.  Although this conformal fixed point doesn't participate in the infrared dynamics, it is clearly ``nearby'' in theory space, and $\Delta_\Phi$ proves to be a useful quantity in tracking the various possible behaviors of the Lifshitz geometry.  In any case, for fixed $\tilde{u}$, $\Delta_\Phi$ can be regarded as a well-defined function of $z>1$.  We plot its behavior in figure~\ref{DoublePhase}.  Recall from section~\ref{CONFORMAL} that $\Delta_\Phi > 2$ on fairly general grounds.%

Figure~\ref{DoublePhase} depicts curves of constant $\tilde{u}$ in $z$-$\Delta_{\Phi}$ space. There is a distinction between two regimes:
 \begin{itemize}
  \item Weak quartic coupling, $\tilde{u} < 1$.  In this regime, $\Delta_{\Phi}$ first increases with $z$, then decreases, with a maximum at $z=z_m$.  When the quartic coupling is weak in this sense, it is possible for a domain wall with emergent conformal symmetry to exist at the same value of $q$ as two different solutions with Lifshitz-like scaling.
  \item Strong quartic coupling, $\tilde{u} > 1$.  This regime is simpler because at every value of $\Delta_{\Phi}$, our analysis leads to only one candidate ground state: an $AdS_4$-to-$AdS_4$ domain wall if $\Delta_{\Phi} > 3$, and an $AdS_4$-to-Lifshitz domain wall if $\Delta_{\Phi} < 3$.
 \end{itemize}

If we specify $m^2$ and $L$, then each point $(z, \Delta_{\Phi})$ in figure~\ref{DoublePhase} that corresponds to a Lifshitz solution can be classified further according to the behaviors of the powers $\beta_{\psi}(1, \pm 1)$.  (Note that with $m^2$ and $L$ fixed, varying $z$ and $\Delta_\Phi$ is equivalent to varying $q$ and $u$.)  The analysis of the $\beta_{\psi}(1, \pm 1)$ proceeds similarly to the analogous analysis of $z$-$m^2L^2$ space in the quadratic case. Since it is tedious and only analytical up to a point, we do not present the details here. The $\beta_{\psi}(1,\pm 1)$ fall into one of five categories, where we have indicated in each case the color of the corresponding region in figure~\ref{DoublePhase}:
\begin{enumerate}
  \item (green) $\beta_{\psi}(1,1)$ and $\beta_{\psi}(1,-1)$ are real and positive. They are associated with irrelevant perturbations, and an $AdS_4$-to-Lifshitz domain wall is probably possible.
  \item (gray) $\beta_{\psi}(1,1)$ and $\beta_{\psi}(1,-1)$ are complex with $\beta_\psi(1,1)=\beta_\psi(1,-1)^*$, and $\Re\beta_\psi > 0$. The two associated perturbations are irrelevant, and a flow to a conformal fixed point in the ultraviolet is likely possible. Such a flow would exhibit damped oscillations in the infrared.
  \item (blue) $\beta_{\psi}(1,1)$ and $\beta_{\psi}(1,-1)$ are real, but one is negative and the other is positive. The negative power is associated with a relevant perturbation and the positive power is associated with an irrelevant perturbation. In general, a flow to a conformal fixed point in the ultraviolet is not possible when one forbids explicitly symmetry breaking deformations of the ultraviolet theory.
  \item (red) $\beta_{\psi}(1,1)$ and $\beta_{\psi}(1,-1)$ are real and negative. They are associated with relevant perturbations, and a flow to a conformal fixed point in the ultraviolet is not possible.
  \item (purple) $\beta_{\psi}(1,1)$ and $\beta_{\psi}(1,-1)$ are complex with $\beta_{\psi}(1,1)=\beta_{\psi}(1,-1)^*$, and $\Re\beta_{\psi} < 0$. The two associated perturbations are relevant, and a flow to a conformal fixed point in the ultraviolet is not possible.
\end{enumerate}

To produce figure~\ref{PhaseDiagram}, we chose $m^2=-2$ and $L=1$. In the white space, there are no Lifshitz solutions. The curve that separates the white space from the colored regions represents the limit $\tilde{u}\rightarrow 0$. At the critical point $(z,\Delta_{\Phi})\approx (1.715, 3.061)$ in the weak-coupling regime, both $\beta_{\psi}(1,1)$ and $\beta_{\psi}(1,-1)$ vanish, and the five colored regions meet. Note from \eno{FinalBFCondition} that $m^2L^2=-2$ always violates the $AdS_2$ BF bound encoded in the inequality \eno{m2zBFCondition}; if $m^2L^2$ had been $-1$, for instance, symmetry-breaking solutions with nonzero $\psi$ probably would not occur in a region of $z$-$\Delta_{\Phi}$ space, approximately where the $AdS_2$ BF bound is satisfied.

At the points $B$, $C$, and $D$, which are displayed in figure~\ref{DoublePhase} as well as figure~\ref{PhaseDiagram}, we have numerically obtained flows to conformal fixed points in the ultraviolet for $m^2=-2$, $L=1$. In figure~\ref{BFlow}, we exhibit the solution that corresponds to point $B$ using the metric convention \eno{AdSAnsatz}. This solution interpolates between two copies of $AdS_4$ and is qualitatively similar to the solution discussed in \cite{Gubser:2008wz}. The UV-to-IR speed-of-light ratio is approximately $2.368$. In figures~\ref{CFlow} and~\ref{DFlow}, we exhibit the solutions that correspond to points $C$ and $D$, respectively, using the metric convention in \eno{BGansatz}. Each interpolates between $AdS_4$ in the ultraviolet and a Lifshitz geometry in the infrared.  Though the solutions at points $C$ and $D$ look similar, they represent qualitatively different behavior in the deep infrared, where $\log r/L_0\rightarrow -\infty$. At point $D$, $\beta_{\psi}(1,1)=\beta_{\psi}(1,-1)^*\approx 3.143+0.910i$: There is a nonzero imaginary part. However, the real part is over three times larger than the imaginary part, and though the solution exhibits oscillations in the infrared, they are so damped that they cannot be seen in the figure. At point $C$, $\beta_{\psi}(1,1)=2.802$ and $\beta_{\psi}(1,-1)=1.245$: Both powers are real and positive, so there are no oscillations in the infrared.

\section{Conclusions}

A prominent feature of quantum field theory is that when an operator that usually is irrelevant or marginal becomes relevant, interesting new dynamics arises: For example, BCS superconductivity and confinement can be understood in these terms.  Here we have a novel example, where the operator in question is the time component, $J_0$, of a vector operator, $J_\mu$, which is conserved in the ultraviolet, but which becomes non-conserved due to condensation of a scalar operator at sufficiently large chemical potential for $J_0$.  In the infrared, if $J_0$ acquires an anomalous dimension large enough to make it irrelevant, relativistic conformal symmetry can be recovered.  If $J_0$ is relevant, then, at least for a broad class of examples typified by the examples in figures~\ref{ColdBH}, \ref{BFlow}, \ref{CFlow}, and \ref{DFlow}, the result is Lifshitz-like scaling in the infrared.

An unexpected feature of flows from $AdS_4$ in the ultraviolet to Lifshitz solutions in the infrared is that the approach to Lifshitz behavior can be oscillatory: in fact, for the positive mass quadratic potential, this appears to be the only possibility.  In field theory, the oscillations presumably represent oscillatory or cyclic approach to the Lifshitz fixed point behavior.  In regions of the bulk geometry where the oscillations are strong, the blackening function, $-g_{tt}/g_{xx}$, can be almost constant over a significant range of values of $g_{xx}$.  An example of this can be seen in figure~\ref{ColdBH}.  Examples with more pronounced shelves can be constructed.  A shelf (nearly constant $-g_{tt}/g_{xx}$) indicates the approximate recovery of an $SO(2,1)$ symmetry over a finite range of energy scales.  Presumably, Green's functions of the dual gauge theory would reflect such an approximate symmetry: in particular, the spectral measure of two-point functions would have its weight concentrated in a momentum-space light-cone with a speed of light determined by $\sqrt{-g_{tt}/g_{xx}}$, over a range of energies corresponding to the extent of the shelf.  An example of this was seen in \cite{Gubser:2008gr} for the case of true emergent conformal symmetry in the infrared.  On the gravity side, one can understand the presence of shelves in $-g_{tt}/g_{xx}$ heuristically as competition between oscillatory behavior and the constraint that $-g_{tt}/g_{xx}$ is a monotonically increasing function of $r$.  This latter constraint follows from \eno{heom}.

A comprehensive study of flows from $AdS_4$ to Lifshitz behavior for the W-shaped quartic potential is clearly an involved task.  There are three dimensionless parameters: $m^2 L^2$, $qL$, and $uL^2$.\footnote{In principle, $\kappa/L$ is another dimensionless parameter.  But $\kappa$ doesn't enter into the equations of motion following from \eno{Lagrangian}, so its value doesn't affect classical solutions.}  The solutions found in figures~\ref{BFlow}-\ref{DFlow} are representative, but to work out the full story, one should investigate to what extent the $AdS_2$ BF bound condition is an accurate guideline to when symmetry breaking solutions exist, and also whether $AdS_4$-to-Lifshitz solutions win out thermodynamically over $AdS_4$-to-$AdS_4$ solutions when they both exist.

We leave open two important questions about stability.  First, are the extremal backgrounds we construct stable against linearized perturbations?  The oscillatory perturbations of Lifshitz solutions have some similarities with scalars that violate of the BF bound in anti-de Sitter space, but it is not clear to us whether they indicate true instabilities of the domain wall solutions.  Second, what is the energetically preferred extremal background at finite chemical potential?  Sometimes---for instance, at large $q$---it is fairly clear that the $AdS_4$-to-$AdS_4$ solutions are indeed the preferred ground state.  But when $\Delta_\Phi$ is only slightly larger than $3$ and the quartic coupling is small, there can be competition between $AdS_4$-to-$AdS_4$ and $AdS_4$-to-Lifshitz domain walls.  It is numerically challenging to ascertain which type of domain wall wins out.  We hope to report on these and related issues in future work.

\section*{Acknowledgements}

We thank A.~Cherman, T.~Cohen, C.~Herzog, S.~Parameswaran, S.~Pufu, M.~Roberts, F.~Rocha, and A.~Yarom for useful discussions.  This work was supported in part by the Department of Energy under Grant No.\ DE-FG02-91ER40671 and by the NSF under award number PHY-0652782.

\clearpage
\bibliographystyle{ssg}
\bibliography{domain}

\begingroup\raggedright\begin{thebibliography}{10}

\bibitem{Gubser:2008px}
S.~S. Gubser, ``{Breaking an Abelian gauge symmetry near a black hole
  horizon},'' {\em Phys. Rev.} {\bf D78} (2008) 065034,
  \href{http://xxx.lanl.gov/abs/0801.2977}{{\tt 0801.2977}}.

\bibitem{Gubser:2005ih}
S.~S. Gubser, ``Phase transitions near black hole horizons,'' {\em Class.
  Quant. Grav.} {\bf 22} (2005) 5121--5144,
  \href{http://xxx.lanl.gov/abs/hep-th/0505189}{{\tt hep-th/0505189}}.

\bibitem{Herzog:2007ij}
C.~P. Herzog, P.~Kovtun, S.~Sachdev, and D.~T. Son, ``{Quantum critical
  transport, duality, and M-theory},'' {\em Phys. Rev.} {\bf D75} (2007)
  085020, \href{http://xxx.lanl.gov/abs/hep-th/0701036}{{\tt hep-th/0701036}}.

\bibitem{Hartnoll:2007ih}
S.~A. Hartnoll, P.~K. Kovtun, M.~Muller, and S.~Sachdev, ``{Theory of the
  Nernst effect near quantum phase transitions in condensed matter, and in
  dyonic black holes},'' {\em Phys. Rev.} {\bf B76} (2007) 144502,
  \href{http://xxx.lanl.gov/abs/0706.3215}{{\tt 0706.3215}}.

\bibitem{Hartnoll:2007ip}
S.~A. Hartnoll and C.~P. Herzog, ``{Ohm's Law at strong coupling: S duality and
  the cyclotron resonance},'' {\em Phys. Rev.} {\bf D76} (2007) 106012,
  \href{http://xxx.lanl.gov/abs/0706.3228}{{\tt 0706.3228}}.

\bibitem{Hartnoll:2008vx}
S.~A. Hartnoll, C.~P. Herzog, and G.~T. Horowitz, ``{Building an AdS/CFT
  superconductor},'' \href{http://xxx.lanl.gov/abs/0803.3295}{{\tt 0803.3295}}.

\bibitem{Gubser:2008wz}
S.~S. Gubser and F.~D. Rocha, ``{The gravity dual to a quantum critical point
  with spontaneous symmetry breaking},''
  \href{http://xxx.lanl.gov/abs/0807.1737}{{\tt 0807.1737}}.

\bibitem{Hartnoll:2008kx}
S.~A. Hartnoll, C.~P. Herzog, and G.~T. Horowitz, ``{Holographic
  Superconductors},'' \href{http://xxx.lanl.gov/abs/0810.1563}{{\tt
  0810.1563}}.

\bibitem{Gubser:2008pf}
S.~S. Gubser and A.~Nellore, ``{Low-temperature behavior of the Abelian Higgs
  model in anti-de Sitter space},''
  \href{http://xxx.lanl.gov/abs/0810.4554}{{\tt 0810.4554}}.

\bibitem{Gubser:2009gp}
S.~S. Gubser, S.~S. Pufu, and F.~D. Rocha, ``{Quantum critical superconductors
  in string theory and M- theory},''
  \href{http://xxx.lanl.gov/abs/0908.0011}{{\tt 0908.0011}}.

\bibitem{Gubser:2009qm}
S.~S. Gubser, C.~P. Herzog, S.~S. Pufu, and T.~Tesileanu, ``{Superconductors
  from Superstrings},'' \href{http://xxx.lanl.gov/abs/0907.3510}{{\tt
  0907.3510}}.

\bibitem{GubserStrings}
S.~S. Gubser, ``Superconducting black holes.''
\newblock Strings 2009 talk, {\tt
  http://strings2009.roma2.infn.it/talks/Gubser\_Strings09.pdf}.

\bibitem{Gauntlett:2009dn}
J.~P. Gauntlett, J.~Sonner, and T.~Wiseman, ``{Holographic superconductivity in
  M-Theory},'' \href{http://xxx.lanl.gov/abs/0907.3796}{{\tt 0907.3796}}.

\bibitem{Gauntlett:2009zw}
J.~P. Gauntlett, S.~Kim, O.~Varela, and D.~Waldram, ``{Consistent
  supersymmetric Kaluza--Klein truncations with massive modes},'' {\em JHEP}
  {\bf 04} (2009) 102, \href{http://xxx.lanl.gov/abs/0901.0676}{{\tt
  0901.0676}}.

\bibitem{Kachru:2008yh}
S.~Kachru, X.~Liu, and M.~Mulligan, ``{Gravity Duals of Lifshitz-like Fixed
  Points},'' {\em Phys. Rev.} {\bf D78} (2008) 106005,
  \href{http://xxx.lanl.gov/abs/0808.1725}{{\tt 0808.1725}}.

\bibitem{Azeyanagi:2009pr}
T.~Azeyanagi, W.~Li, and T.~Takayanagi, ``{On String Theory Duals of
  Lifshitz-like Fixed Points},'' {\em JHEP} {\bf 06} (2009) 084,
  \href{http://xxx.lanl.gov/abs/0905.0688}{{\tt 0905.0688}}.

\bibitem{Li:2009pf}
W.~Li, T.~Nishioka, and T.~Takayanagi, ``{Some No-go Theorems for String Duals
  of Non-relativistic Lifshitz-like Theories},''
  \href{http://xxx.lanl.gov/abs/0908.0363}{{\tt 0908.0363}}.

\bibitem{Mack:1975je}
G.~Mack, ``{All Unitary Ray Representations of the Conformal Group SU(2,2) with
  Positive Energy},'' {\em Commun. Math. Phys.} {\bf 55} (1977) 1.

\bibitem{Grinstein:2008qk}
B.~Grinstein, K.~A. Intriligator, and I.~Z. Rothstein, ``{Comments on
  Unparticles},'' {\em Phys. Lett.} {\bf B662} (2008) 367--374,
  \href{http://xxx.lanl.gov/abs/0801.1140}{{\tt 0801.1140}}.

\bibitem{Bertoldi:2009vn}
G.~Bertoldi, B.~A. Burrington, and A.~Peet, ``{Black Holes in asymptotically
  Lifshitz spacetimes with arbitrary critical exponent},''
  \href{http://xxx.lanl.gov/abs/0905.3183}{{\tt 0905.3183}}.

\bibitem{Bertoldi:2009dt}
G.~Bertoldi, B.~A. Burrington, and A.~W. Peet, ``{Thermodynamics of black
  branes in asymptotically Lifshitz spacetimes},''
  \href{http://xxx.lanl.gov/abs/0907.4755}{{\tt 0907.4755}}.

\bibitem{Breitenlohner:1982bm}
P.~Breitenlohner and D.~Z. Freedman, ``{Positive Energy in anti-De Sitter
  Backgrounds and Gauged Extended Supergravity},'' {\em Phys. Lett.} {\bf B115}
  (1982) 197.

\bibitem{Breitenlohner:1982jf}
P.~Breitenlohner and D.~Z. Freedman, ``{Stability in Gauged Extended
  Supergravity},'' {\em Ann. Phys.} {\bf 144} (1982) 249.

\bibitem{Denef:2009tp}
F.~Denef and S.~A. Hartnoll, ``{Landscape of superconducting membranes},'' {\em
  Phys. Rev.} {\bf D79} (2009) 126008,
  \href{http://xxx.lanl.gov/abs/0901.1160}{{\tt 0901.1160}}.

\bibitem{Faulkner:2009wj}
T.~Faulkner, H.~Liu, J.~McGreevy, and D.~Vegh, ``{Emergent quantum criticality,
  Fermi surfaces, and AdS2},'' \href{http://xxx.lanl.gov/abs/0907.2694}{{\tt
  0907.2694}}.

\bibitem{Gubser:2008gr}
S.~S. Gubser, ``{Time warps},'' \href{http://xxx.lanl.gov/abs/0812.5107}{{\tt
  0812.5107}}.

\end{thebibliography}\endgroup
\end{document}